\documentclass[twocolumn]{aastex631} 

\usepackage{amsmath} 
\usepackage{placeins} 

\usepackage{xcolor}

\shorttitle{Complete R-Process Survey}
\shortauthors{Kuske, Arcones, Reichert}

\begin{document}

\title{Complete survey of r-process conditions: \\the (un-)robustness of the r-process(-es)}

\author[0009-0005-5121-7343]{Jan Kuske}
\affiliation{Institut für Kernphysik, Technische Universität Darmstadt, Schlossgartenstr. 2, Darmstadt 64289, Germany}

\author[0000-0002-6995-3032]{Almudena Arcones}
\affiliation{Institut für Kernphysik, Technische Universität Darmstadt, Schlossgartenstr. 2, Darmstadt 64289, Germany}
\affiliation{GSI Helmholtzzentrum für Schwerionenforschung GmbH, Planckstr. 1, Darmstadt 64291, Germany}
\affiliation{Max-Planck-Institut für Kernphysik, Saupfercheckweg 1, 69117 Heidelberg, Germany}

\author[0000-0001-6653-7538]{Moritz Reichert}
\affiliation{Departament d'Astronomia i Astrof\'{\i}sica, Universitat de Val\`encia, C/Dr. Moliner, 50, 46100 Burjassot (Val\`encia), Spain}



\begin{abstract}
Heavy elements are synthesized by the r-process in neutron star mergers and potentially in rare supernovae linked to strong magnetic fields. Expensive hydrodynamic simulations of these extreme environments are usually post-processed to calculate the nucleosynthesis. In contrast, here we follow a site-independent approach based on three key parameters: electron fraction, entropy, and expansion timescale. Our model reproduces the results based on hydrodynamic simulations. Moreover, the 120\,000 astrophysical conditions analyzed allow us to systematically and generally explore the astrophysical conditions of the r-process, also beyond those found in current simulations. Our results show that a wide range of conditions produce very similar abundance patterns explaining the observed robustness of the r-process between the second and third peak. Furthermore, we cannot find a single condition that produces the full r-process from first to third peak. Instead, a superposition of at least two or three conditions or components is required to reproduce the typical r-process pattern as observed in the solar system and very old stars. The different final abundances are grouped into eight nucleosynthesis clusters, which can be used to select representative conditions for comparisons to observations and investigations of the nuclear physics input.
\end{abstract}



\section{Introduction} 
\label{sec:intro}

In the last decade, great progress has been made in understanding the r-process and its contribution to the production of heavy elements beyond iron \citep{Horowitz2019_RProcessNucleosynthesisConnecting, Cowan2021_OriginHeaviestElements}. In the r-process, neutrons are captured faster than the beta decays of the involved nuclei. Therefore, this process moves far from stability reaching unknown neutron-rich nuclei and requires extreme explosive conditions. Here we focus on the astrophysical conditions required to have a successful r-process and explore them in a site-independent way building on previous works \citep{Meyer1997_SurveyRProcessModels, Freiburghaus1999_AstrophysicalRprocessComparison, Arnould2007_RprocessStellarNucleosynthesis, Lippuner2015_RprocessLanthanideProduction}. 

At least one site for the r-process has been identified with the observation of a kilonova \citep{Abbott2017_MultimessengerObservationsBinary,Drout2017_LightCurvesNeutron,Kilpatrick2017_ElectromagneticEvidenceThat} after the gravitational wave detection GW170817 following a neutron star merger (NSM) \citep{Abbott2017_GW170817ObservationGravitational}. Moreover, galactic chemical evolution and observation of heavy elements in the atmospheres of very old stars indicate that at least one additional site had to produce r-process elements at early times \cite[see e.g.,][]{Cote2019_NeutronStarMergers, Molero2023_OriginNeutronCapture}. One possibility is core-collapse supernovae with strong magnetic fields and rotation, namely magneto-rotational supernovae (MRSN). The early prompt ejecta are triggered by the magnetic pressure and stay neutron-rich, allowing for the r-process to build heavy elements \citep{Nishimura2006_RProcessNucleosynthesisMagnetohydrodynamic,Winteler2012_MagnetorotationallyDrivenSupernovae, Mosta2018_RprocessNucleosynthesisThreedimensional, Reichert2022_MagnetorotationalSupernovaeNucleosynthetic,Zha2024_NucleosynthesisInnermostEjecta}. Moreover, the late phase of such explosions can lead to a collapsar or magnetar with also favorable conditions for the r-process \cite[see e.g.,][]{Siegel2019_CollapsarsMajorSource,Patel2025_DirectEvidenceRprocess}.

Given the diversity of proposed scenarios and the possibility of finding new r-process sites, we follow the approach of exploring all possible conditions without a link to a given astrophysical site. This has the advantage of exploring not only the conditions found so far, but also new potential ones. Moreover, our study can reproduce the nucleosynthesis based on trajectories from hydrodynamic simulations. With this broader study, we have found important constraints for the r-process. Our results show that the robustness observed between the second and third r-process peaks does not necessarily come from always having the same conditions, but from a large number of conditions producing the same abundance pattern. Another important conclusion discussed also in previous works \cite[see e.g.,][]{Qian2007_WhereOhWhere, Hansen2014_HowManyNucleosynthesis} is the need of at least two contributions to explain the full r-process pattern from the first to the third peak. Furthermore, our goal is also to identify nucleosynthesis groups and representative trajectories that can be used in the future to explore the impact of nuclear physics and compare with observations, following the strategy for the weak r-process from \cite{Bliss2018_SurveyAstrophysicalConditions, Bliss2020_NuclearPhysicsUncertainties, Psaltis2022_ConstrainingNucleosynthesisNeutrinodriven, Psaltis2024_NeutrinodrivenOutflowsElemental}. 

This paper is structured as follows: In Sect.~\ref{sec:model_nuc}, we provide an overview of our model, the nuclear reaction network, and its inputs. Sect.~\ref{sec:results} gives a detailed overview of the initial conditions and the final abundances. Sect.~\ref{sec:comparisonToHydroTrajectories} compares the results of our model with those of trajectories from hydrodynamical simulations. In Sect.~\ref{sec:comparisonToObservations}, we compare with observed abundance patterns. We summarize our findings and conclude in Sect.~\ref{sec:conclusions}.

\section{Model and nucleosynthesis} 
\label{sec:model_nuc}

The r-process can be studied by post-processing trajectories from simulations but also by parametric studies. The advantage of those is that they can systematically cover a broad range of conditions. A disadvantage is the smooth evolution of thermodynamic properties such as density and temperature that cannot include shocks or similar non-monotonic conditions self consistently (see Sect.~\ref{sec:comparisonToHydroTrajectories}). Nevertheless, parametric studies have predictive power as the assumption of homologous expansion is generally valid in r-process ejecta.

\subsection{Parametric model}
\label{sec:model}

\begin{figure}
    \centering
    \includegraphics[width=\columnwidth]{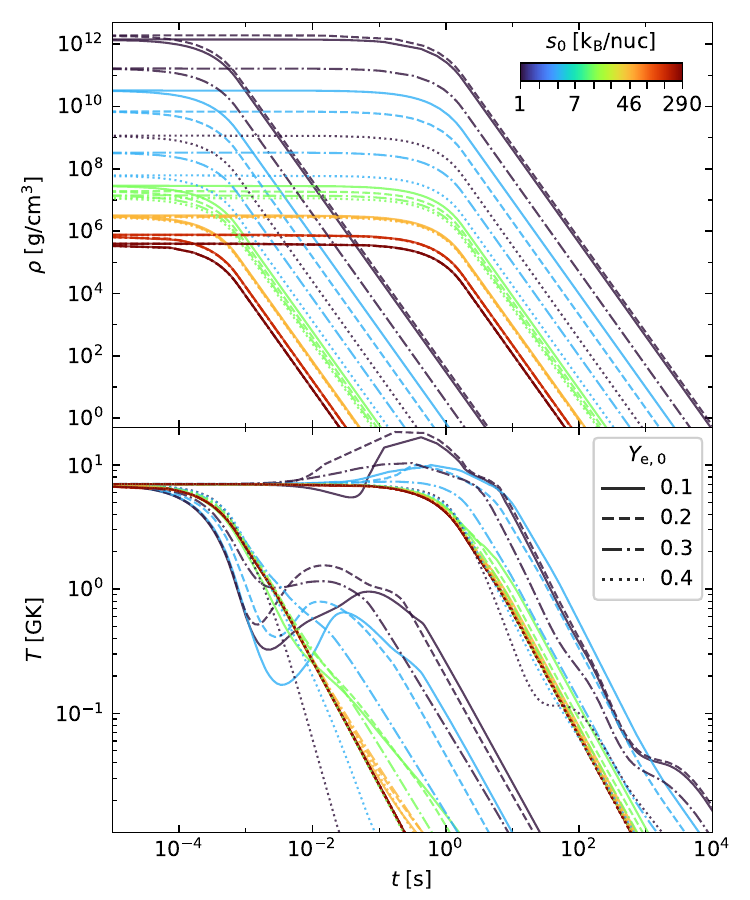}
    \caption{Density and temperature evolution for a selection of conditions: Two extreme expansion timescales with several initial entropies and electron fractions are shown. The density profile is described by Eq.~(\ref{eq:densityParameterization}), while the temperature is calculated from the Timmes EOS, including the entropy increase from nuclear heating. This leads to non-monotonic behaviors for low initial entropies.}
    \label{fig:mainGridDensAndTemp}
\end{figure}

Similarly to previous studies \cite[e.g.,][]{Qian1996_NucleosynthesisNeutrinodrivenWinds, Hoffman1997_ModelIndependentRprocess, Hoffman1997_NucleosynthesisNeutrinodrivenWinds, Freiburghaus1999_AstrophysicalRprocessComparison, Lippuner2015_RprocessLanthanideProduction}, we choose as parameters the initial specific entropy $s_\mathrm{0}$, the initial electron fraction $Y_\mathrm{e,0}$, and the expansion timescale $\tau$. These quantities are given at the initial temperature of our model, $T_0=7$~GK, and the initial density $\rho_0$ (Sect.~\ref{sec:ini_cond}) is derived assuming nuclear statistical equilibrium (NSE) and using the equation of state (EOS) from \cite{Timmes1999_AccuracyConsistencySpeed}. 
The temperature evolution is based on an adiabatic expansion and includes nuclear energy generation. We have explored different possibilities for the density evolution \cite[e.g.,][]{Meyer2002_RProcessExcessNeutrons} and found the best agreement with most hydrodynamic trajectories when using the expansion of \cite{Lippuner2015_RprocessLanthanideProduction}. We follow their density profile which is made up of an initial exponential decrease followed by a homologous expansion:
\begin{equation}
    \rho(t) = \rho_0 \begin{cases} \exp{\big(-\frac{t}{\tau}\big)} & \text { if } t \leq 3 \tau \\ \big(\frac{3\tau}{e t}\big)^3 & \text { if } t \geq 3 \tau\end{cases}~,
    \label{eq:densityParameterization}
\end{equation}
where $\tau$ is the expansion timescale. The density and temperature evolution of selected conditions is shown in Fig.~\ref{fig:mainGridDensAndTemp} for two extreme expansion timescales and various initial electron fractions and entropies.

 Within the model described above, we have varied the input parameters as follows: 100 linearly spaced values between 0.005 and 0.5 for $Y_\mathrm{e,0}$; 100 logarithmically spaced values between 1 and 290~$\mathrm{k_B/nuc}$ for $s_\mathrm{0}$; 12 logarithmically spaced values between 0.25 and 600~ms for $\tau$. These variations correspond to a total of 120\,000 individual trajectories that cover all possible astrophysical neutron-rich conditions found in current simulations of neutron star mergers, neutrino-driven, and magneto-rotational-driven supernovae. Moreover, other conditions are explored beyond current simulations, including some interesting regions of the parameter space as well as others that are far from typical conditions found in r-process sites and probably not realized in any astrophysical environment. Compared to previous studies (Table~\ref{tab:parameterRanges}), we include a broader range of conditions than in \citet{Hoffman1997_ModelIndependentRprocess,Hoffman1997_NucleosynthesisNeutrinodrivenWinds} [H+97]\footnote{H+97 investigated the minimum entropy required for an r-process to occur for a given $Y_\mathrm{e}$ and $\tau$.}, \citet{Meyer1997_SurveyRProcessModels} [MB97], and \citet{Freiburghaus1999_AstrophysicalRprocessComparison} [F+99]. \citet{Lippuner2015_RprocessLanthanideProduction} [LR15] focused on neutron star merger conditions. Their thermodynamical setup and range of conditions are similar to ours, however, we have a finer parameter grid and different nuclear physics input (see Sect.~\ref{sec:netw}). Furthermore, we focus our analysis on different quantities, e.g., the agreement with trajectories from hydrodynamical simulations (Sect.~\ref{sec:comparisonToHydroTrajectories}) and observed stellar abundances (Sect.~\ref{sec:comparisonToObservations}). 
 
\begin{table}
\caption{Comparison of parameter spaces.}
\label{tab:parameterRanges}
\centering
    \begin{tabular}{l c c c}
\hline
\hline
Reference & $Y_\mathrm{e,0}$ &  $s_\mathrm{0}$                &  $\tau$  \\
          &           &  $\mathrm{[k_B/nuc]}$ &  [ms] \\
\hline 
    This work & 0.005-0.5 (100) & 1-290   (100) & 0.25-600  (12) \\
    H+97      & 0.2-0.5   (11)  & 28-532  (1)  & 5-250     (5) \\
    MB97      & 0.1-0.5   (41)  & 50-500  (10)  & 1-1000    (12) \\
    F+99      & 0.29-0.49 (11)  & 3-390   (40)  & 50-150    (2) \\
    LR15      & 0.01-0.5  (17)  & 1-100   (17)  & 0.1-500   (17) \\
\hline        
\end{tabular}
\tablecomments{The numbers in brackets specify the number of different values for each parameter.}
\end{table}

\subsection{Nucleosynthesis network}
\label{sec:netw}

We use the open-source nuclear reaction network code \textsc{WinNet} \citep{Reichert2023_NuclearReactionNetwork}, considering 7583 nuclei up to $^{337}\mathrm{Og}$. Experimental data are used where available, most other theoretical nuclear properties are based on the FRDM2012 mass model \citep{Moller2017_NuclearGroundstateMasses}: The neutron capture rates are calculated with the Hauser-Feshbach code \textsc{Talys} \citep{Koning2023_TALYSModelingNuclear} and the photodisintegration rates are obtained from detailed balance. The beta decay rates are from \citet{Moller2019_NuclearPropertiesAstrophysical}, fission rates from \citet{Panov2005_CalculationsFissionRates,Panov2010_NeutroninducedAstrophysicalReaction,Khuyagbaatar2020_SpontaneousFissionHalflives} using the fission barriers from \citet{Moller2015_FissionBarriersEnd}, and fragment distributions from \citet{Mumpower2020_PrimaryFissionFragment}. The remaining reactions are taken from the \textsc{REACLIB} database (\citealt{Cyburt2010_JinaReaclibDatabase}, version from September 18th 2020). For $T>0.01\,\mathrm{GK}$, temperature and density-dependent theoretical weak rates of \citet{Langanke2001_RateTablesWeak}, \citet{Oda1994_RateTablesWeak}, \citet{Fuller1985_StellarWeakInteraction}, \citet{Pruet2003_EstimatesStellarWeak}, and \citet{Suzuki2016_ElectroncaptureVdecayRates} have been used as described in \cite{Reichert2023_NuclearReactionNetwork}. For $T<0.01\,\mathrm{GK}$, these theoretical weak rates are replaced by the experimental decay rates contained in the \textsc{REACLIB} database. To simplify our model, we chose not to explicitly include neutrino reactions, as this would introduce additional parameters. By directly varying $Y_\mathrm{e,0}$ in our model, we still capture their impact on the electron fraction.

We have implemented a new parametric mode in \textsc{WinNet}. This mode allows to specify a function for the time evolution of density $\rho(t)$, in addition to the initial values of electron fraction $Y_\mathrm{e,0}$ and the specific entropy per nucleon $s_\mathrm{0}$, as well as either the initial temperature $T_0$ (used here) or density $\rho_0$.  At each time step, the temperature is derived from the EOS, taking into account changes in density, composition, and entropy. The latter is updated taking into account the energy generation from nuclear reactions \cite[see][for more details]{Reichert2023_NuclearReactionNetwork}. The heating description assumes that neutrinos can freely escape from the medium. However, at high densities, this approximation would lead to an artificial and rapid decline in entropy through the loss of neutrinos by electron/positron captures, as well as beta decays. To avoid this problem, nuclear heating is activated only once the density has dropped below $10^{11}~\mathrm{g\,cm^{-3}}$. Notice that the Timmes EOS may become uncertain at high densities ($\rho \gtrsim 10^{11}~\mathrm{g\,cm^{-3}}$). When the initial entropy and electron fraction are low, the initial density is high and nuclear heating can lead to increases in temperature above $10\,\mathrm{GK}$ (see Fig.~\ref{fig:mainGridDensAndTemp}), at which point \textsc{WinNet} switches from network to NSE evolution mode. It switches back to the network once the temperature has dropped below $7\,\mathrm{GK}$.

\section{Results of r-process survey}
\label{sec:results}

\subsection{Initial NSE conditions}
\label{sec:ini_cond}

\begin{figure}
    \centering
    \includegraphics[width=\columnwidth]{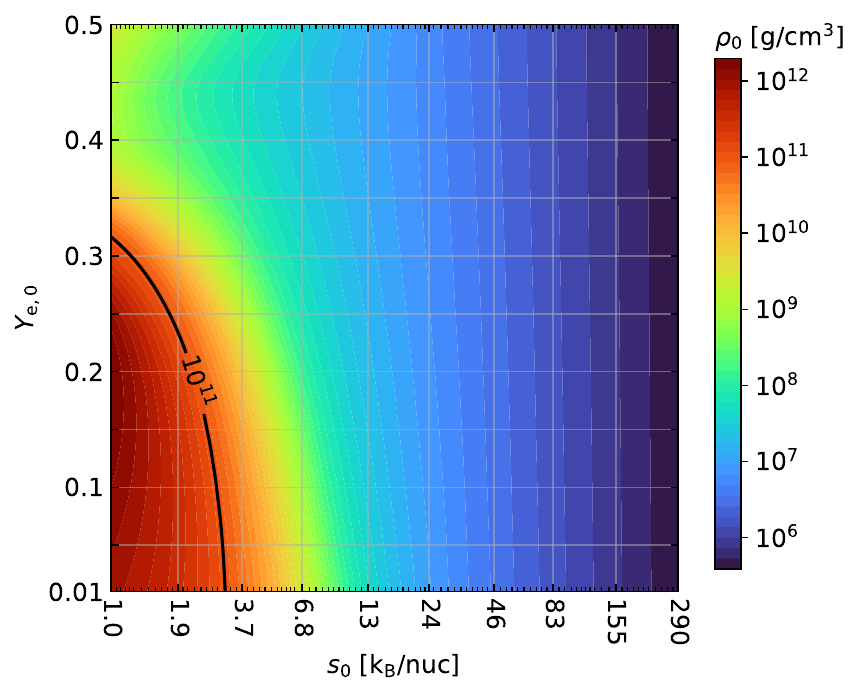}
    \caption{Initial densities as a function of entropy $s_\mathrm{0}$ and electron fraction $Y_\mathrm{e,0}$ as calculated from NSE at a temperature of $7\,\mathrm{GK}$. The black line marks initial densities of $10^{11}\,\mathrm{g\,cm^{-3}}$, above which the  EOS is less precise as it does not include nuclear contributions.}
    \label{fig:initDensity}
\end{figure}

\begin{figure*}
    \centering
    \includegraphics[width=\textwidth]{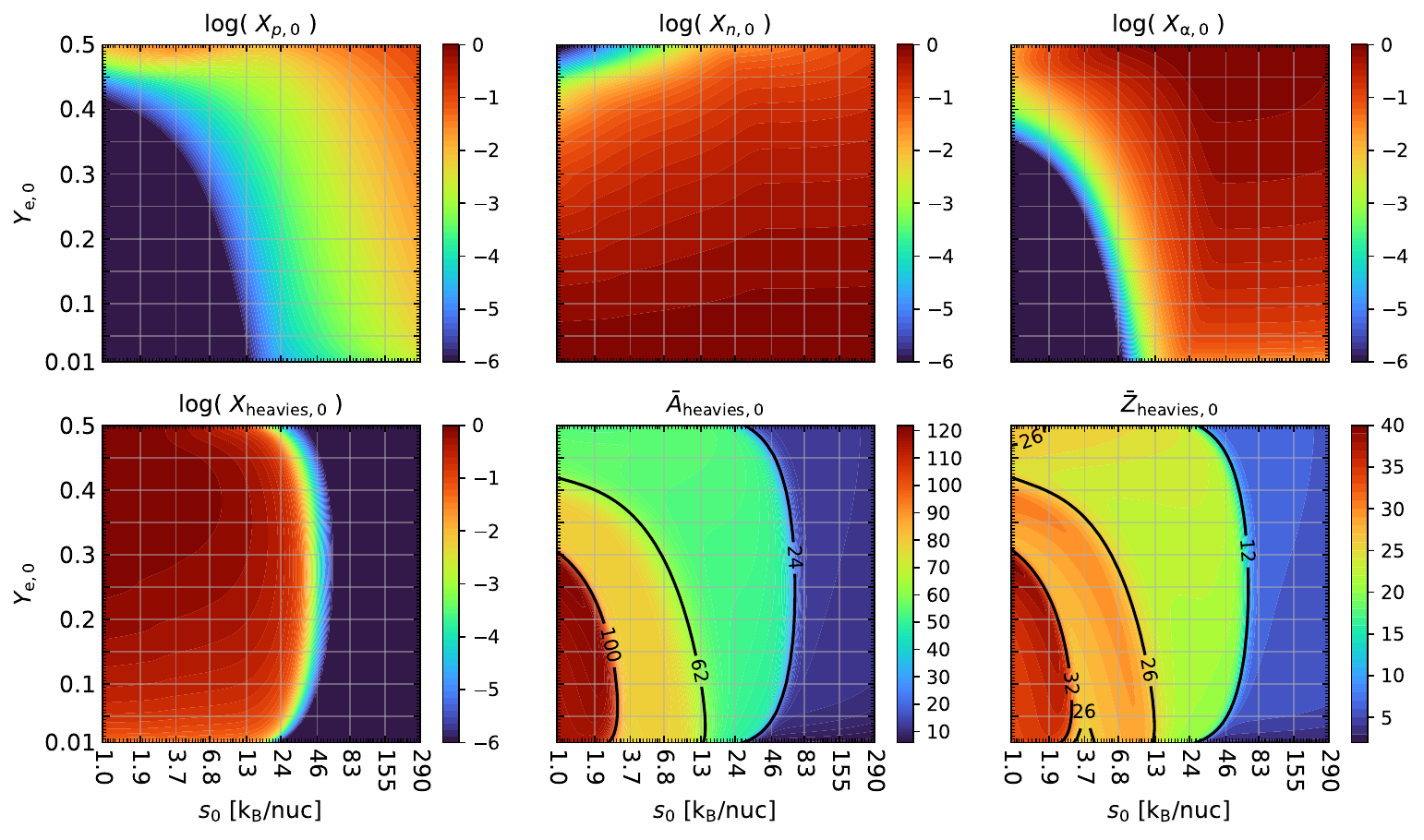}
    \caption{Initial NSE mass fractions of protons, neutrons, alphas, and heavies ($A>4$), as well as the average mass and proton numbers of heavies.}
    \label{fig:initMassFractions}
\end{figure*}

Figure~\ref{fig:initDensity} shows the initial densities from NSE at $7\,\mathrm{GK}$ as a function of $Y_\mathrm{e,0}$ and $s_\mathrm{0}$. The solid black line marks the conditions at which the initial density is $10^{11}~\mathrm{g\,cm^{-3}}$. The initial mass fractions of protons, neutrons, alphas, and heavy nuclei ($A>4$) as well as the average mass and proton numbers of heavy elements are shown in Fig.~\ref{fig:initMassFractions}. In order to understand these initial conditions, we distinguish two regimes: The ion-dominated regime at $s_\mathrm{0} \lesssim 30\,\mathrm{k_B/nuc}$ and $Y_\mathrm{e}\lesssim0.35$ and the radiation-dominated regime elsewhere.

In the ion-dominated regime, heavy nuclei and free neutrons predominate the abundances with almost no protons, light nuclei (i.e., deuteron, triton, and $^3\mathrm{He}$), and alphas. The entropy resides almost exclusively in ions. The ion entropy in the Timmes EOS is treated via the Sackur-Tetrode equation \citep{Sackur1911_AnwendungKinetischenTheorie,Tetrode1912_ChemischeKonstanteGase} as
\begin{align}
    s_\mathrm{ion}
    &= \left( \frac{P_\mathrm{ion}}{\rho} + E_\mathrm{ion}\right)\frac{1}{T}-\frac{\eta_\mathrm{ion} k_B N_\mathrm{ion}}{\rho}\\
\label{eq:sIon}
    &= \frac{k_B N_A}{\bar A} \Bigg(\frac{5}{2}-\underbrace{\ln{\left(\frac{N_A h^3}{(2\pi m_u)^{3/2}}\frac{\rho}{(k_BT)^{3/2}\bar A^{5/2}} \right)}}_{\substack{=\eta_\mathrm{ion}}}\Bigg)\,,
\end{align}
where $P_\mathrm{ion}$, $E_\mathrm{ion}$, $N_\mathrm{ion}$, and $\eta_\mathrm{ion}$ denote the pressure, energy, number, and degeneracy parameter of ions, $N_A$ is the Avogadro number, $m_u$ the atomic mass unit, while $k_B$ and $h$ are the Boltzmann and Planck constants. Here only one representative nucleus is considered, such that $N_\mathrm{ion}/\rho=N_A/\bar A$, where
\begin{align}
\label{eq:Abar}
    \bar A=\frac{\sum_i A_i Y_i}{\sum_i Y_i}=\frac{\sum_i X_i}{\sum_i Y_i}=\frac{1}{\sum_i Y_i}
\end{align} is the average mass number of all nuclei from neutrons and protons to the heaviest nuclei. Furthermore, $A_i$, $Y_i$, and $X_i$ are the mass number, abundance and mass fraction of nucleus $i$, respectively. Rearranging Eq.~(\ref{eq:sIon}) for $\rho$ and assuming $s\approx s_\mathrm{ion}$ gives the relation
\begin{align}
\label{eq:approxIonEOS}
    \rho\propto T^{3/2}\bar A^{5/2}\exp{(-s\bar A)}\,.
\end{align}
Therefore, the density in the ion-dominated regime increases exponentially with lower entropy and reaches its maximum for $\bar A=2.5/s$. In our grid this maximum is reached for $s_\mathrm{0}=1\,\mathrm{k_B/nuc}$ and $Y_\mathrm{e,0}=0.185$, where $\bar A=2.6$ (indicating that the above assumption $s\approx s_\mathrm{ion}$ is quite accurate). Note that the lower center panel of Fig.~\ref{fig:initMassFractions} depicts the initial average mass number of heavies $\bar A_\mathrm{ heavies,0}$, which differs from $\bar A$, since it does not consider free nucleons and light nuclei in the sums of Eq.~(\ref{eq:Abar}). The values of $\bar A_\mathrm{ heavies,0}$ and $\bar Z_\mathrm{ heavies,0}$ approximately follow the density, however, they show a non-monotonic behavior around densities of $10^{11}\,\mathrm{g\,cm^{-3}}$.

At high entropy, the system consists mainly of free nucleons and alpha particles with a minor contribution of heavy nuclei. In this radiation-dominated environment, the density fulfills the relation $\rho\propto T^3/s$. Free neutrons dominate the initial abundances at $Y_\mathrm{e,0}<0.25$, otherwise alpha particles have the largest mass fraction. Free protons appear only for the highest electron fractions and entropies in our parameter space.

\subsection{Final abundances}
\label{sec:fin_ab}

Before we explore the full parameter space, we highlight here the general dependencies of the final abundances on the three nucleosynthesis parameters. Heavy r-process elements beyond the second peak ($A\gtrsim130$) are produced for low electron fractions, high entropies, and fast expansion timescales, as extensively discussed in previous works \cite[see e.g.,][and references therein]{Hoffman1997_ModelIndependentRprocess,Hoffman1997_NucleosynthesisNeutrinodrivenWinds,Meyer1997_SurveyRProcessModels,Arnould2007_RprocessStellarNucleosynthesis,Wanajo2018_PhysicalConditionsRprocess,Lippuner2015_RprocessLanthanideProduction}. These conditions increase the neutron-to-seed ratio ($Y_\mathrm{n}/Y_\mathrm{seed}$) that together with the average mass number of the seeds ($\bar{A}_\mathrm{seed}$) will determine which heavy elements can be produced \citep{Hoffman1997_NucleosynthesisNeutrinodrivenWinds}:
\begin{align}
    \label{eq:AFinApprox}
    A_\mathrm{fin}\approx \bar{A}_\mathrm{seed}+\frac{Y_\mathrm{n}}{Y_\mathrm{seed}}\,.
\end{align}

\begin{figure}
    \centering
    \includegraphics[width=\columnwidth]{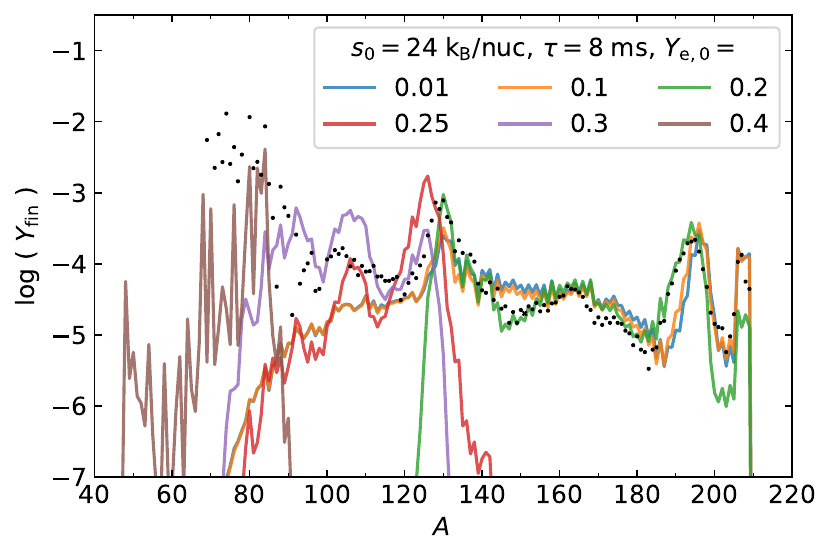}
    \caption{Final abundances for different initial electron fractions. The solar r-process residuals from \cite{Sneden2008_NeutronCaptureElementsEarly} are shown as black dots, normalized to $Y(A=151)=2\cdot10^{-5}$.}
    \label{fig:constYeSTauAbund}
\end{figure}

The neutron-to-seed ratio is directly linked to the initial electron fraction. The final abundances (i.e., after 1\,Gyr) for different electron fractions are shown in Fig.~\ref{fig:constYeSTauAbund} for a representative entropy ($s_\mathrm{0}=24\,\mathrm{k_B/nuc}$) and expansion timescale ($\tau=8\,\mathrm{ms}$). Initial electron fractions of $0.4$ lead to small neutron-to-seed ratios, therefore, only nuclei up to the first peak ($A\approx80$) are produced. When conditions are slightly more neutron-rich ($Y_\mathrm{e,0}=0.3$), the r-process path can overcome the first peak but it eventually stops at $N=82$ shell closure, i.e., the second peak ($A\approx130$). Below a threshold value of around $Y_\mathrm{e,0}=0.25$ the third r-process peak ($A\approx195$) can be produced, as also found in previous studies \citep[e.g.,][]{Lippuner2015_RprocessLanthanideProduction, Freiburghaus1999_AstrophysicalRprocessComparison}. For very low electron fractions, fission cycling occurs and leads to a robust final abundance pattern that is independent of the exact conditions \citep{Korobkin2012_AstrophysicalRobustnessNeutron, Mendoza-Temis2015_NuclearRobustnessProcess}. In this case, the abundances are characterized by fission products contributing at $A<120$, high Pb and Bi after alpha decays, and a shifted third peak due to neutron captures after freeze-out \cite[see e.g.,][]{Arcones2011_DynamicalRprocessStudies}.

Higher entropies correspond to less seed nuclei, which also increase the neutron-to-seed ratio. The expansion timescale sets the time interval in which alpha particles can fuse into seed nuclei before the start of the r-process, i.e., around 3\,GK. Shorter expansion timescales increase the neutron-to-seed ratio. Therefore, high entropies and fast expansions also facilitate the production of heavy nuclei.

\begin{figure*}
    \centering
    \includegraphics[width=\textwidth]{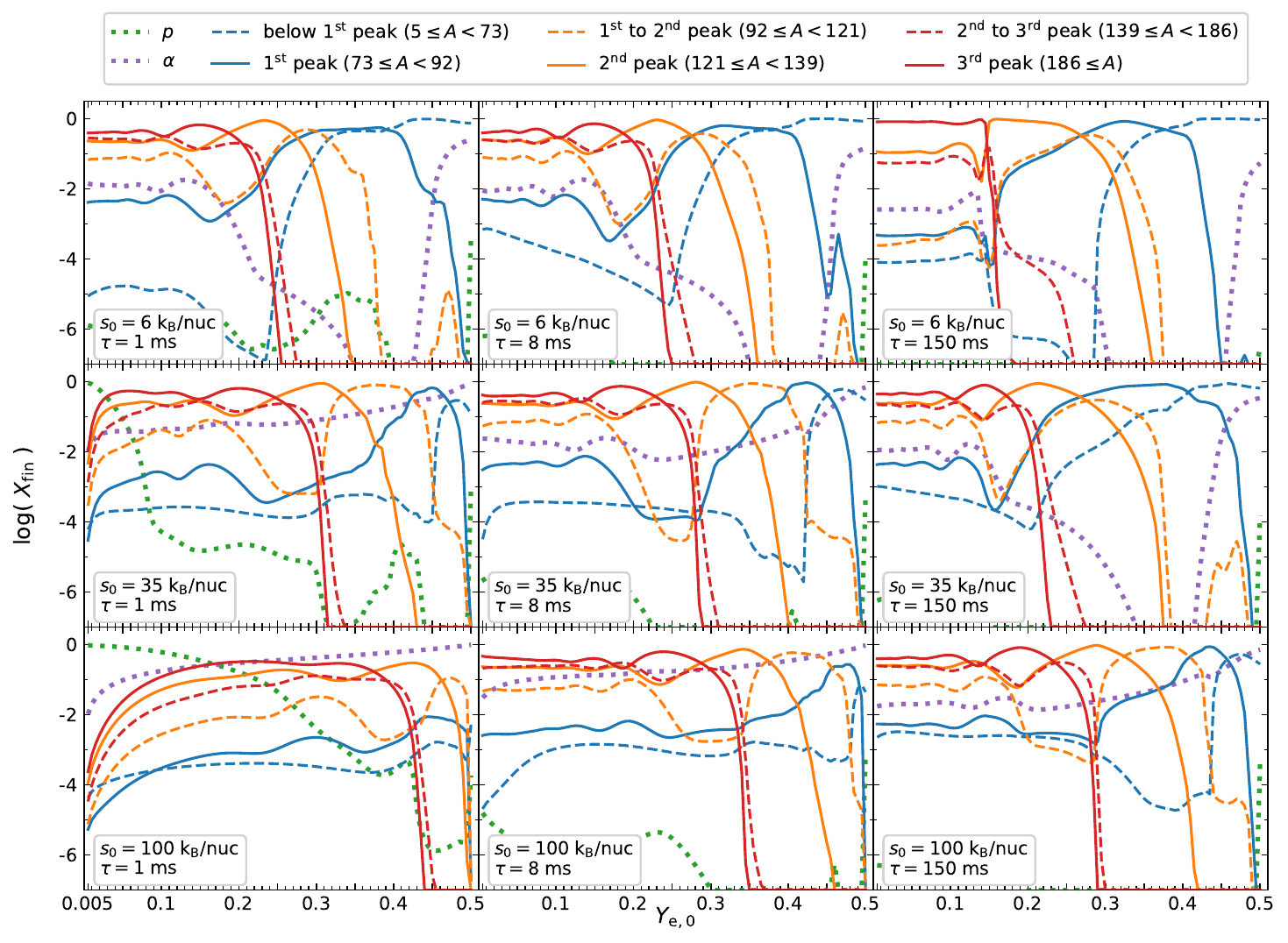}
    \caption{Final mass fractions of protons, alphas, and different groups of nuclei. Their dependence on the initial electron fraction $Y_\mathrm{e,0}$ is shown for a variety of initial entropies $s_\mathrm{0}$ (rows) and expansion timescales $\tau$ (columns).}
    \label{fig:2DMultiple}
\end{figure*}

A more complete overview can be found in Fig.~\ref{fig:2DMultiple}: Each panel shows final mass fractions as a function of the electron fraction for various entropies ($s_\mathrm{0}$=~6, 35, and 100~$\mathrm{k_B/nuc}$ in rows) and expansion timescales ($\tau$=~1, 8, and 150~$\mathrm{ms}$ in columns). Mass fractions are shown for protons, alphas, and representative groups of nuclei around and between the peaks.

Generally, the first peak (blue solid line) is formed at $Y_\mathrm{e,0}$ between approximately 0.3 and 0.4. At higher $Y_\mathrm{e,0}$, iron-group nuclei dominate (blue dashed line). Second peak nuclei (orange solid line) are created roughly for $Y_\mathrm{e,0}\lesssim0.35$ and the third peak (red solid line) is usually produced for $Y_\mathrm{e,0}\lesssim0.25$. While the third peak is always accompanied by the second peak, it is rarely co-produced with the first peak. In all conditions with $s_0 \gtrsim 2\,\mathrm{k_B/nuc}$, the threshold $Y_\mathrm{e,0}$ for the production of second and third peaks increases with higher initial entropies and shorter expansion timescales, as the neutron-to-seed ratio increases.

In the cases where the third peak is efficiently produced, fission cycling plays an important role on the final abundances. As discussed in \mbox{\citet{Eichler2019_ProbingProductionActinides,Holmbeck2019_ActiniderichActinidepoorRprocessenhanced}}, higher or lower production of the third peak and of actinides depends on when the r-process freeze-out occurs compared to the ongoing fission cycle. This is visible as oscillations in the final mass fractions as a function of $Y_\mathrm{e,0}$ (see mass fraction for second and third peaks in Fig.~\ref{fig:2DMultiple}, e.g., panel for  $s_\mathrm{0}=6\,\mathrm{k_B/nuc}$ and $\tau=8\,\mathrm{ms}$).

Alpha particles contribute to the final abundances in two cases. First, an alpha-rich freeze-out occurs for $0.4\lesssim Y_\mathrm{e,0}\lesssim0.5$ and not too low entropies ($s_\mathrm{0}\gtrsim2\,\mathrm{k_B/nuc}$). This has been extensively studied in neutrino-driven supernovae (see e.g.~\cite{Bliss2018_SurveyAstrophysicalConditions, Arcones2013_NeutrinodrivenWindSimulations} for an overview). Below $Y_\mathrm{e,0}\lesssim0.25$ alphas are co-produced with third peak nuclei by alpha decays.  For each expansion timescale, there exists a critical entropy value, above which there are always significant amounts of alphas in the final abundances. This critical value is around $20\,\mathrm{k_B/nuc}$ for very fast expansions, which is similar to the entropy threshold for the existence of alphas in the initial abundances shown in Fig.~\ref{fig:initMassFractions}. For slower expansions, there is increasingly more time for alphas to form seed nuclei, thus the critical entropy increases to about $100\,\mathrm{k_B/nuc}$ for the slowest expansions of our grid.

In conditions with extremely large neutron-to-seed ratios (i.e., low electron fraction, high entropy, and fast expansion) many neutrons are left after the r-process freeze-out. These neutrons eventually decay and result in high proton fractions (green dotted line) in the final abundances.

\subsection{Nucleosynthesis groups}
\label{sec:groups}

\begin{figure*}
    \centering
    \includegraphics[width=\textwidth]{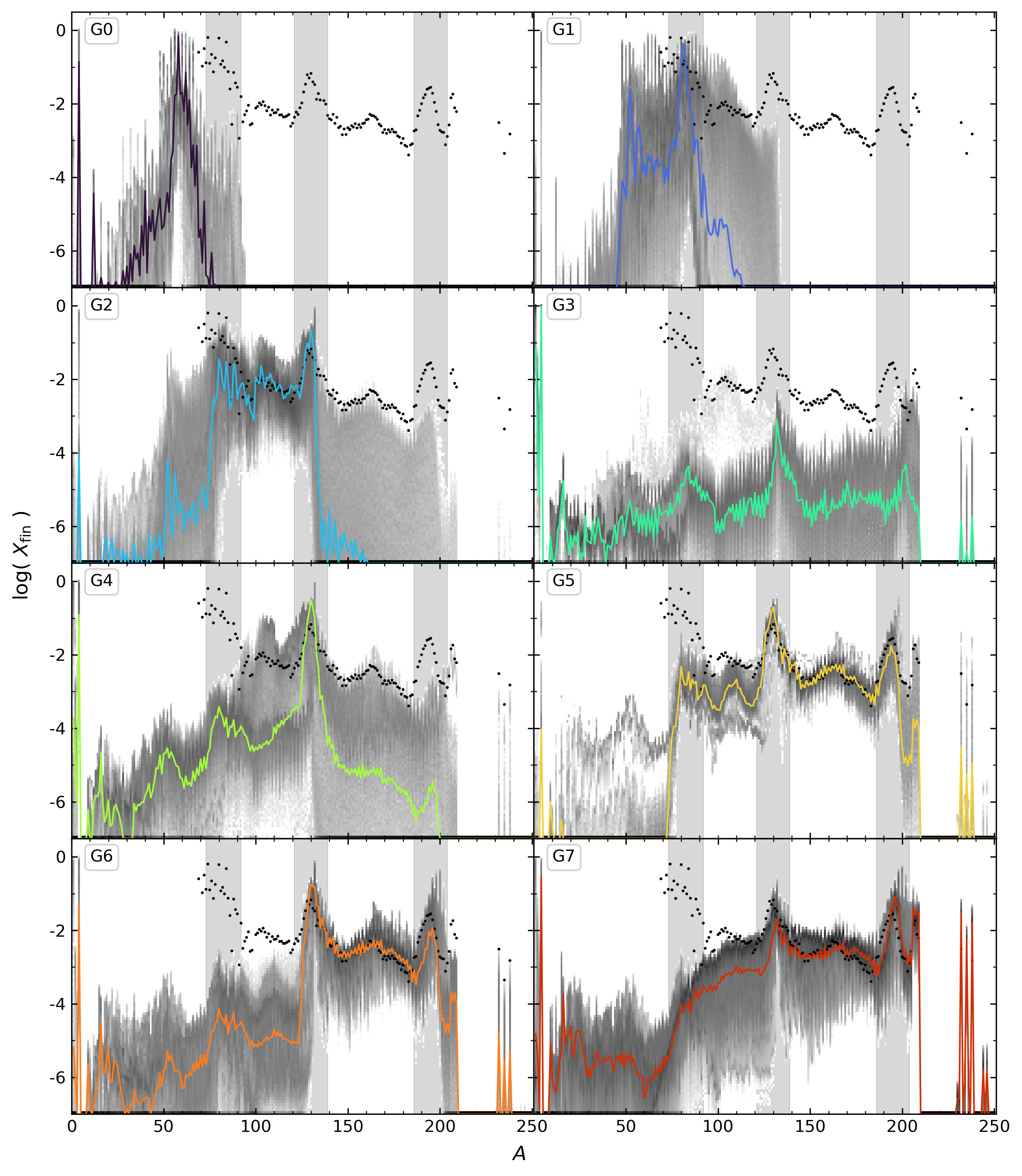}
    \caption{Final mass fractions as a function of mass number for the eight nucleosynthesis groups summarized in Table~\ref{tab:groups}. The colored line shows the most representative condition of each group, the other conditions are shown as shades of gray. The solar r-process residuals from \cite{Sneden2008_NeutronCaptureElementsEarly} are shown as black dots, normalized such that $X(A=151)=2\cdot10^{-3}$. The mass intervals used for determining the groups correspond to the gray bands.}
    \label{fig:clustersByPeaksAbundances}
\end{figure*}

\begin{figure*}
    \centering
    \includegraphics[angle=0, width=\textwidth]{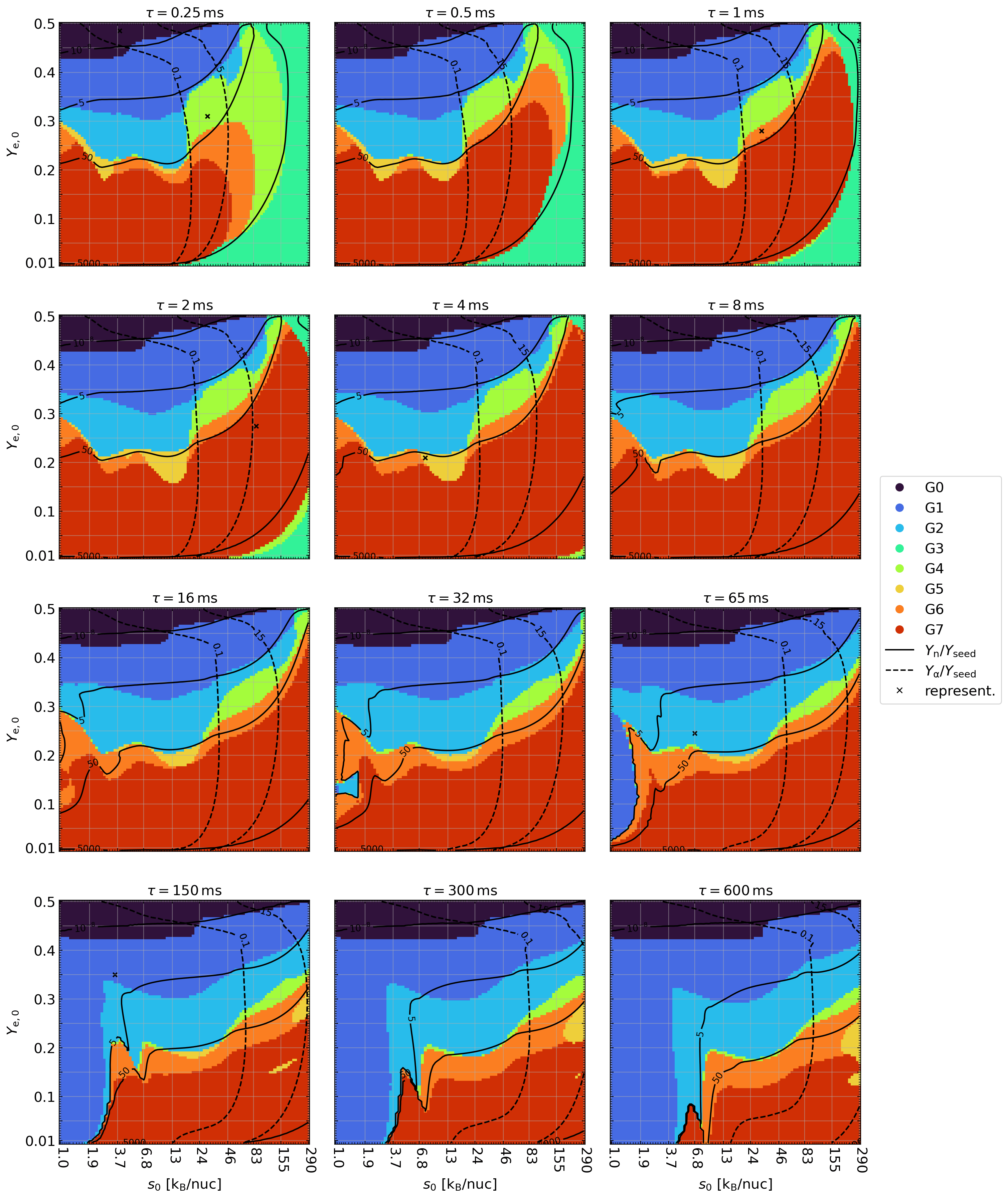}
    \caption{Regions in the parameter space that produce the eight nucleosynthesis groups G0 to G7 from Fig.~\ref{fig:clustersByPeaksAbundances}. The solid  (dashed) lines display the indicated neutron-to-seed (alpha-to-seed) ratios at $T=3\,\mathrm{GK}$. Crosses mark the most representative condition of each group.}
    \label{fig:clustersByPeaks}
\end{figure*}

\begin{figure}
    \centering
    \includegraphics[width=\columnwidth]{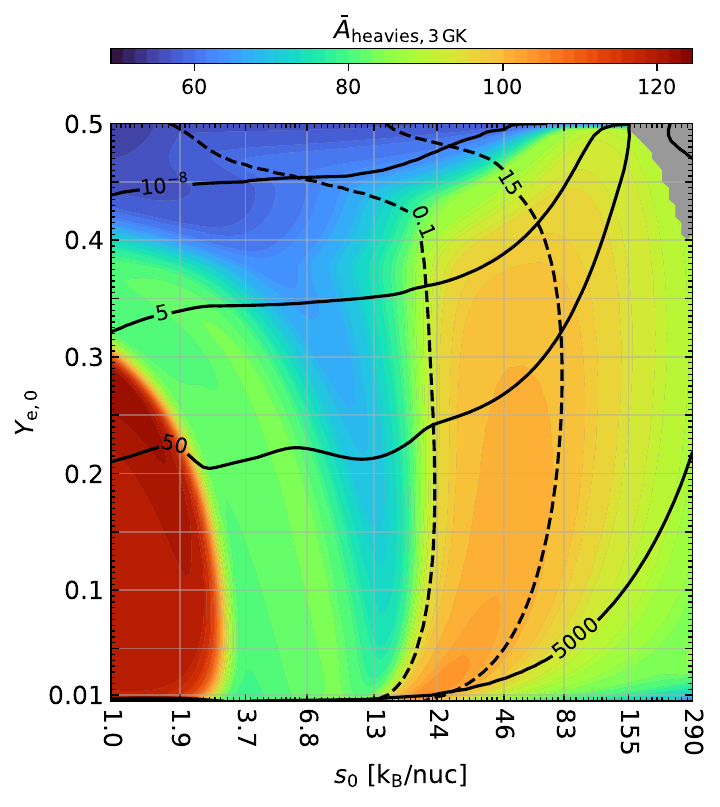}
    \caption{Average mass number of heavy nuclei ($A>4$) at 3\,GK for $\tau=2\,\mathrm{ms}$. The solid and dashed lines are the neutron-to-seed and alpha-to-seed ratios at 3\,GK, respectively. Below entropies of about 13\,$\mathrm{k_B/nuc}$, the values are almost identical to the initial NSE values of $\bar{A}_\mathrm{heavies}$ in Fig.~\ref{fig:initMassFractions}.}
    \label{fig:AbarHeavies3GK}
\end{figure}

We group our 120\,000 calculations by their final abundances based on the production of r-process peaks. We use the following mass number intervals to define the first, second, and third peaks, respectively: $73\leq A\leq 91$, $121\leq A\leq 138$, and $186\leq A\leq 203$. A peak is counted as `produced' if $\sum_A X(A)>10^{-2}$ (cf. Fig.~\ref{fig:clustersByPeaksAbundances}).
We further subdivide the groups as summarized in Table~\ref{tab:groups}. In addition to the grouping criteria, the table provides the number of conditions per group ($N_\mathrm{conditions}$) and mean value, upper and lower $1\sigma$ intervals for the initial electron fraction, entropy, expansion timescale, and average mass number of heavy nuclei. The last column corresponds to the mean value of the differences between each set of final mass fractions and the group average\footnote{The average mass fraction pattern of a group is defined as: $ X_\mathrm{average}(A)=\sum X_i(A)/N_\mathrm{conditions}$.}. To quantify the differences between two sets of mass fractions $X_i(A)$ and $X_j(A)$, we introduce the following difference averaged over all mass numbers:
\begin{align}
    \label{eq:massFractionDistance}
    d_{i,j}=\frac{1}{A_\mathrm{max}}\sum_{A=1}^{A_\mathrm{max}}|\log \big(X_i(A)\big)-\log\big(X_j(A)\big)|\,,
\end{align}
where we use $A_\mathrm{max}=250$ as the maximum mass number and we consider only mass fractions above $10^{-7}$. Therefore, in Table~\ref{tab:groups}, $d_{i,\mathrm{average}}$ gives the mass fraction difference between each group member ($i$) and the average pattern. The mean value $\langle d_{i,\mathrm{average}}\rangle$ reflects whether the abundance patterns of a group are robust or deviate significantly from each other. For each group, a solid line in Fig.~\ref{fig:clustersByPeaksAbundances} shows the abundance pattern of the most representative conditions, i.e., the trajectory ($i$) for which $d_{i,\mathrm{average}}$ is minimal. All other conditions are gray-shaded, where darker colors indicate a higher density of conditions.

Figure~\ref{fig:clustersByPeaks} displays the conditions in our parameter space that produce these groups. To understand the final abundances and the link of the groups to the conditions, one has to consider three key quantities: neutron-to-seed ratio (solid lines mark $Y_\mathrm{n}/Y_\mathrm{seed} =10^{-8},5,50,5000$), alpha-to-seed ratio (dashed lines mark $Y_\mathrm{\alpha}/Y_\mathrm{seed}=0.1,15$.), and average mass number of seed nuclei at 3\,GK, cf. Eq.~(\ref{eq:AFinApprox}). The latter is shown in Fig.~\ref{fig:AbarHeavies3GK} for $\tau=2\,\mathrm{ms}$. We find that the heaviest seeds with $\bar{A}_\mathrm{seed}\gtrsim 100$ are produced when the abundances of alphas and seeds at 3\,GK are of the same order of magnitude, i.e., $Y_\mathrm{\alpha}/Y_\mathrm{seed}\approx1$ (regions between dashed lines in Figs.~\ref{fig:clustersByPeaks} and \ref{fig:AbarHeavies3GK}). For such conditions, there were enough initial alphas for an effective alpha-process \citep{Woosley1992_AlphaProcessRProcess}, but not too many to result in large final alpha abundances. Notice that the ion-dominated regime (i.e., low $s_0$ and low $Y_{e,0}$) marks a second region in our parameter space in which $\bar{A}_\mathrm{seed}$ at 3\,GK can be as high as 120 (red region in Fig.~\ref{fig:AbarHeavies3GK}). Here, heavy seeds are already produced in NSE (cf. Fig.~\ref{fig:initMassFractions}).

\begin{table*}
    \centering
    \caption{Nucleosynthesis groups with their representative and average properties.}
    \begin{tabular}{lrrrrrrrrr}
    \hline
    \hline
    Group: r-Process peaks      & $N_\mathrm{cond}$ & $Y_\mathrm{e,0}^\mathrm{repr}$ & $s_\mathrm{0}^\mathrm{repr}$ & $\tau^\mathrm{repr}$ & $\langle Y_\mathrm{e,0}\rangle$ & $\langle s_\mathrm{0}\rangle_\mathrm{log}$ & $\langle\tau\rangle_\mathrm{log}$ & $\langle \bar A_\mathrm{heavies} \rangle$ & $\langle d_{i,\mathrm{average}}\rangle$ \\
    \, & \, & \, & $\mathrm{[k_B/nuc]}$ & $\mathrm{[ms]}$ & \, & $\mathrm{[k_B/nuc]}$ & $\mathrm{[ms]}$ & \, & \, \\
    \hline
    G0: none (Fe/Ni)            & 10488 & 0.485 &   3.90 &   0.25 & $0.47^{+0.03}_{-0.03}$ &    $5.91^{+16.09}_{-4.26}~$ &  $20.04^{+279.96}_{-19.04}$ &   $58.06^{+2.22}_{-1.82}~$ & $0.26^{+0.06}_{-0.06}$ \\
    G1: 1st                     & 27736 & 0.350 &   3.50 & 150.00 & $0.36^{+0.08}_{-0.05}$ &    $8.93^{+37.07}_{-7.28}~$ &  $23.27^{+276.73}_{-22.27}$ &  $75.38^{+11.75}_{-11.55}$ & $0.50^{+0.19}_{-0.12}$ \\
    G2: 1st + 2nd               & 17341 & 0.245 &   6.75 &  65.00 & $0.28^{+0.05}_{-0.05}$ &   $11.84^{+50.16}_{-9.09}~$ &  $23.06^{+276.94}_{-22.06}$ & $103.32^{+13.83}_{-13.88}$ & $0.49^{+0.09}_{-0.13}$ \\
    G3: none ($n$/$p$/$\alpha$) &  4505 & 0.465 & 290.00 &   1.00 & $0.21^{+0.24}_{-0.18}$ & $171.71^{+88.29}_{-61.71}~$ & $0.52^{+0.48}_{-0.27}~\,\;$ & $116.15^{+45.09}_{-42.85}$ & $0.54^{+0.18}_{-0.22}$ \\
    G4: 2nd                     &  5324 & 0.310 &  29.00 &   0.25 & $0.32^{+0.07}_{-0.06}$ &  $66.98^{+78.02}_{-37.98}~$ &    $2.92^{+29.08}_{-2.67}~$ &  $126.28^{+6.65}_{-10.81}$ & $0.81^{+0.17}_{-0.22}$ \\
    G5: 1st + 2nd + 3rd         &  1697 & 0.210 &   7.75 &   4.00 & $0.21^{+0.02}_{-0.02}$ &  $14.41^{+180.59}_{-11.31}$ &    $5.87^{+144.13}_{-5.37}$ & $138.98^{+14.64}_{-11.74}$ & $0.40^{+0.36}_{-0.15}$ \\
    G6: 2nd + 3rd (ac-poor)     &  8712 & 0.280 &  31.00 &   1.00 & $0.23^{+0.06}_{-0.07}$ &  $20.54^{+99.46}_{-18.32}~$ &  $17.69^{+282.31}_{-17.19}$ & $150.42^{+13.04}_{-12.21}$ & $0.56^{+0.15}_{-0.14}$ \\
    G7: 2nd + 3rd (ac-rich)     & 44197 & 0.275 &  88.00 &   2.00 & $0.12^{+0.08}_{-0.08}$ &  $23.88^{+131.12}_{-20.78}$ &    $8.04^{+56.96}_{-7.04}~$ &  $165.27^{+10.27}_{-9.91}$ & $0.45^{+0.11}_{-0.15}$ \\
    \hline
    \end{tabular}
    \label{tab:groups}
\end{table*}

The group G0 is produced for $Y_\mathrm{e,0}\approx0.5$, low to intermediate entropies, and preferably long expansion timescales. In these conditions, the final abundances reflect mainly those of NSE, which consist almost exclusively of iron-group nuclei without alphas or neutrons. This also leads to only small abundance variations between the conditions of the group. These conditions are typically found in  CCSN.

For slightly more neutron-rich conditions ($10^{-8}\lesssim Y_\mathrm{n}/Y_\mathrm{seed}\lesssim5$ some neutron captures are possible, resulting in group G1. Due to the low neutron fluxes, the $N=50$ shell closure cannot be overcome, and only the first r-process peak is produced. These conditions are mostly reached for $Y_\mathrm{e,0}\approx0.4$, but also for low electron fractions when expansion timescales are long and initial entropies small. The necessary conditions for this `weak r-process' are relatively common and found in almost all considered models (see Table~\ref{tab:trajectoryOverview} and Fig.~\ref{fig:YeSTauMapping}).

The group G2 also contains `weak r-process' conditions that, however, produce the first and second peaks. For $Y_\mathrm{e,0}\approx 0.3$ the intermediate neutron-to-seed ratios of $5\lesssim Y_\mathrm{n}/Y_\mathrm{seed}\lesssim50$ are not enough to overcome the $N=82$ shell closure. The astrophysical conditions are found in neutrino-driven wind ejecta of NSM (for $\tau\leq16\,\mathrm{ms}$), as well as in NSM-DISK and MRSN ejecta (for $\tau\geq8\,\mathrm{ms}$).

In extreme conditions with very high entropies and fast expansion timescales, extremely high neutron-to-seed ratios above 5000 are reached (G3). These allow for the production of the heaviest r-process nuclei, including actinides. However, their abundances are low due to a lack of seed nuclei. After r-process freeze-out, the abundances are dominated by alphas (for $Y_\mathrm{e,0}>0.25$) or neutrons (for $Y_\mathrm{e,0}<0.25$). The high entropies lead to strong variations between the abundance patterns and to a shift of the third peak towards higher mass numbers compared to solar. These conditions could potentially be found in magnetar giant flares \citep{Patel2025_DirectEvidenceRprocess, Patel2025_RProcessNucleosynthesisRadioactively}.

The group G4 is a high entropy edge case of group G2, which produces a strong second peak. Alpha-to-seed ratios at 3\,GK close to unity lead to heavy seed nuclei with $\bar{A}_\mathrm{seed}\approx100$ (Fig.~\ref{fig:AbarHeavies3GK}). These conditions result in small amounts of nuclei beyond the second peak even with intermediate $Y_\mathrm{n}/Y_\mathrm{seed}\approx10$. During this process, the first peak nuclei are depleted, but the third peak cannot be reached. The abundances in this group are highly sensitive to small changes in the conditions, as the large value of $\langle d_\mathrm{i,average}\rangle$ in Table~\ref{tab:groups} shows. Similar conditions are found in shock-heated NSM trajectories with $\tau=1-4\,\mathrm{ms}$, e.g., in \citet{Bovard2017_RprocessNucleosynthesisMatter,Jacobi2023_EffectsNuclearMatter}.

Only few conditions (1.4\% of our parameter space) produce a full r-process pattern with all three peaks (G5). However, nuclei below the second peak are underproduced compared to solar. For most expansion timescales these conditions occur around $Y_\mathrm{e,0}\approx0.2$ and two distinct entropy intervals centered around 3 and 15\,$\mathrm{k_B/nuc}$. Very specific conditions ($Y_\mathrm{n}/Y_\mathrm{seed}\approx50$ and $70\lesssim\bar{A}_\mathrm{seed}\lesssim80$) are required to produce the third peak without depleting the first. Conditions with $s_0 \lesssim 15\,\mathrm{k_B/nuc}$ are commonly produced in many NSM simulations in combination with other conditions. There are also extreme conditions not found in simulations that produce all three peaks with long expansion timescales and very high entropies.

For neutron-to-seed ratios around 50, the second and third peaks are produced (G6). Actinides are only produced for higher neutron to-seed ratios (G7). This group contains the largest number of conditions (cf. large red region in Fig.~\ref{fig:clustersByPeaks}). Moreover, the r-process patterns in this group are robust due to fission cycling. In both groups, the strong neutron fluxes move matter to heavy nuclei, producing the third peak and reducing the first peak. The necessary conditions are commonly found in most of the investigated hydrodynamical models (Sect.~\ref{sec:comparisonToHydroTrajectories}).

The typical r-process pattern with three peaks, as found in metal-poor stars \citep[e.g.,][]{Sneden2003_ExtremelyMetalpoorNeutron, Roederer2022_RprocessAllianceNearly} or the solar r-residuals \citep{Sneden2008_NeutronCaptureElementsEarly,Prantzos2020_ChemicalEvolutionRotating}, is not produced by any of the investigated conditions (see Sect.~\ref{sec:comparisonToObservations}). Even in the small number of conditions that produce all three peaks, the agreement with solar is poor for the first to second peak elements.
This agrees with the observational evidence for (at least) two r-process components in metal-poor stars \cite[e.g.,][]{Qian2001_ModelAbundancesMetalpoor,Qian2007_WhereOhWhere,Hansen2014_HowManyNucleosynthesis}. 

\section{Comparison to hydrodynamic simulation trajectories}
\label{sec:comparisonToHydroTrajectories}

\begin{figure*}
    \centering
    \includegraphics[width=\textwidth]{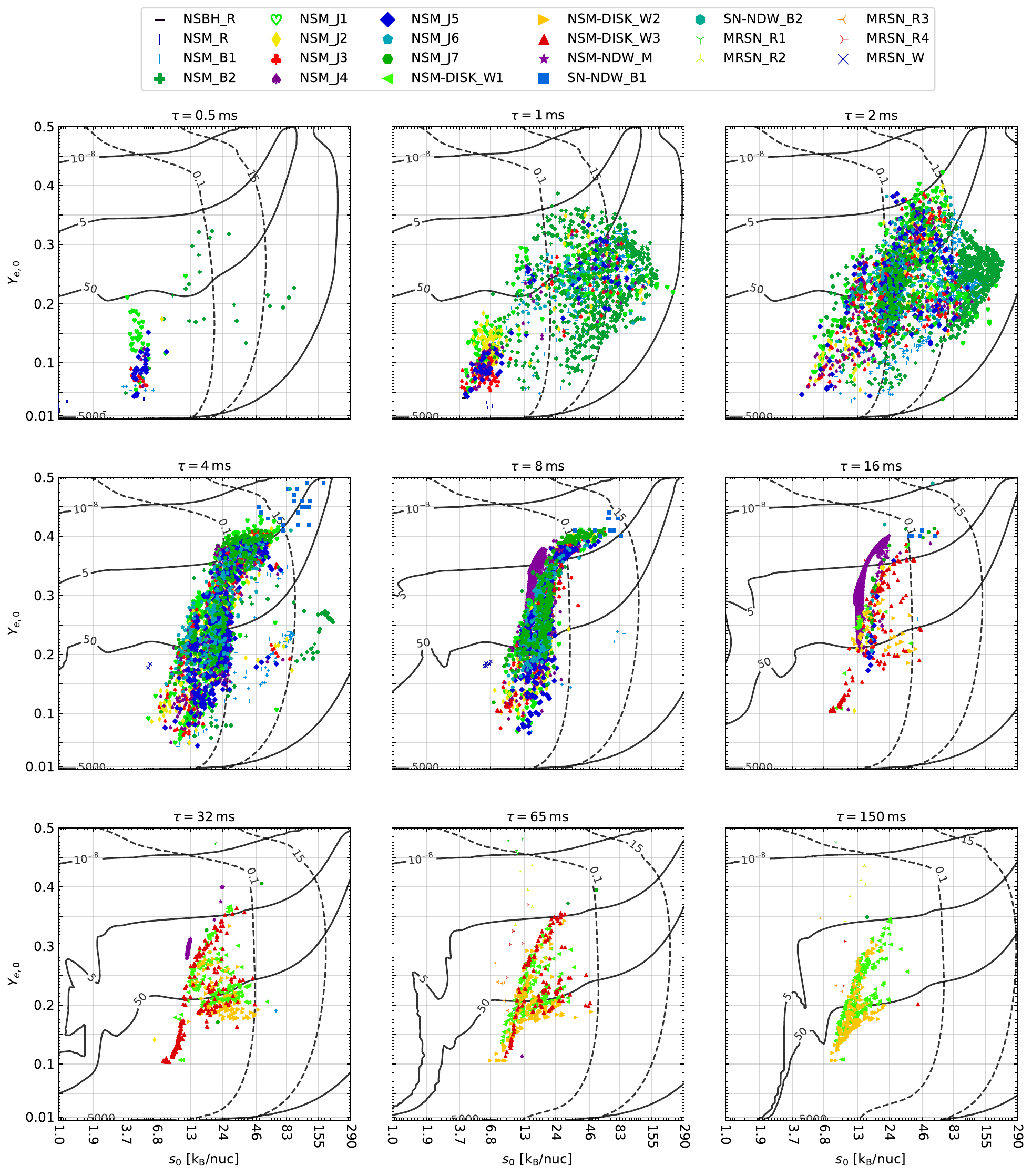}
    \caption{Mapping of trajectories from different astrophysical simulations, summarized in Table~\ref{tab:trajectoryOverview}, into our model space. The values for $Y_\mathrm{e,0}$ and $s_\mathrm{0}$ were interpolated to the last time at which the tracer reaches a temperature of 7\,GK; the values of $\tau$ were extracted from fitting the density profiles to Eq.~(\ref{eq:densityParameterization}). The solid (dashed) lines correspond to the neutron-to-seed (alpha-to-seed) ratio at 3\,GK.}
    \label{fig:YeSTauMapping}
\end{figure*}

We show here the agreement of the parametric density profiles with Lagrangian tracer particles from hydrodynamic simulations of several astrophysical environments summarized in Table~\ref{tab:trajectoryOverview}. For each tracer, the best matching model parameters are determined by interpolating $Y_\mathrm{e,0}$ and $s_\mathrm{0}$ to 7\,GK, and by fitting Eq.~(\ref{eq:densityParameterization}) to the density profile of the tracer for $\tau$. The resulting parameters are shown in Fig.~\ref{fig:YeSTauMapping}.

\begin{figure}
    \centering
    \includegraphics[width=\columnwidth]{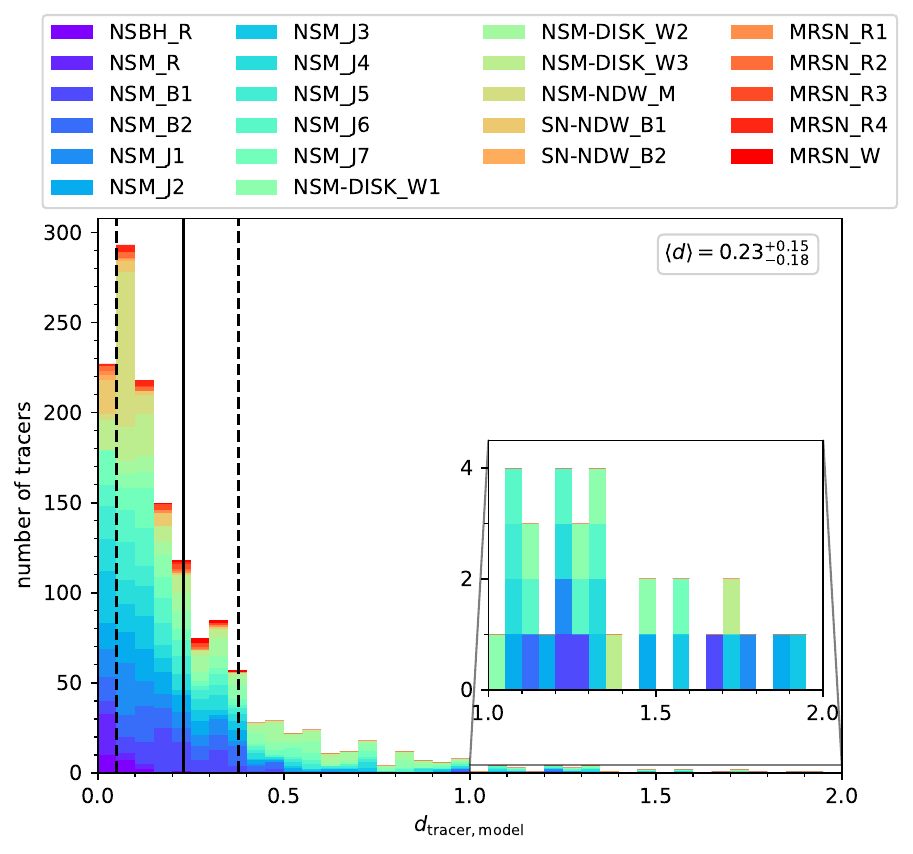}
    \caption{Average mass fraction difference between the final abundances of trajectories and our model, as defined in Eq.~(\ref{eq:massFractionDistance}).  The colors represent the different hydrodynamical models. The solid line is the mean, while the left and right dashed lines mark the lower and upper 1$\sigma$ intervals.}
    \label{fig:diffHistogram_fitted}
\end{figure}

The parametric approach is able to reproduce the nucleosynthesis for most tracers of all considered astrophysical scenarios. We perform nucleosynthesis calculations for up to 100 randomly selected trajectories per model and compare their final mass fractions to those of parametric trajectories using Eq.~(\ref{eq:massFractionDistance}). The results are shown as a histogram in Fig.~\ref{fig:diffHistogram_fitted}. The low mean value of the average mass fraction difference $\langle d_{\mathrm{traj.},\mathrm{model}} \rangle= 0.23^{+0.15}_{-0.18}$ verifies that for most trajectories the predicted final mass fractions differ by less than a quarter order of magnitude. However, there are some outliers with larger differences, which are caused either by non-monotonic density profiles (e.g., shocks) or by sensitive regions of the parameter region between two groups (cf. Sect.~\ref{sec:groups}). In the latter, slight deviations in the hydrodynamics can lead to significant differences in the abundances.

\begin{table*}
    \caption{Hydrodynamical models used for comparison.}
    \label{tab:trajectoryOverview}
    \centering
    \begin{tabular}{llllrrrr}
    \hline
    \hline
    Model & Ref. & Original name & $N_\mathrm{tracers}$ & $\langle Y_\mathrm{e,0}\rangle$ & $\langle s_\mathrm{0}\rangle_\mathrm{log}$ & $\langle\tau\rangle_\mathrm{log}$ & $\langle d_{i, \mathrm{model}}\rangle$\\
    & & & & & $\mathrm{[k_B/nuc]}$ & $\mathrm{[ms]}$ & \\
    \hline
    NSBH\_R      & 1, 2, 3 & ns1.4-bh5.0         &          20 & $0.034^{+0.006}_{-0.018}$ & $ 0.03^{+ 0.49}_{- 0.02}~$ & $  0.32^{+ 0.11}_{- 0.09}~$ & $0.06^{+0.02}_{-0.04}$ \\
    NSM\_R       & 1, 2, 3 & ns1.0-ns1.0         &          30 & $0.032^{+0.011}_{-0.011}$ & $ 0.04^{+ 2.09}_{- 0.03}~$ & $  0.39^{+ 0.19}_{- 0.14}~$ & $0.04^{+0.03}_{-0.03}$ \\
    NSM\_B1      & 4       & LS220-M1.25         &  100 (1102) & $0.226^{+0.068}_{-0.061}$ & $49.29^{+42.34}_{-24.16}$  & $  1.96^{+ 0.51}_{- 0.47}~$ & $0.25^{+0.09}_{-0.14}$ \\
    NSM\_B2      & 4       & SFHO-M1.25          &  100 (1634) & $0.240^{+0.055}_{-0.068}$ & $65.65^{+78.98}_{-36.56}$  & $  1.64^{+ 0.86}_{- 0.49}~$ & $0.22^{+0.15}_{-0.15}$ \\
    NSM\_J1      & 5, 6    & LS220$^\dag$        &  100 (1383) & $0.236^{+0.065}_{-0.069}$ & $20.11^{+10.37}_{- 6.84}$  & $  3.98^{+ 2.25}_{- 1.60}~$ & $0.19^{+0.12}_{-0.14}$ \\
    NSM\_J2      & 5, 6    & LS255$^\dag$        &   100 (940) & $0.245^{+0.107}_{-0.101}$ & $19.22^{+ 9.73}_{- 6.56}~$ & $  4.53^{+ 3.60}_{- 2.31}~$ & $0.25^{+0.13}_{-0.19}$ \\
    NSM\_J3      & 5, 6    & $(m^*K)_\mathrm{S}$ &   100 (992) & $0.240^{+0.113}_{-0.095}$ & $20.11^{+17.22}_{- 7.83}$  & $  3.69^{+ 4.39}_{- 1.79}~$ & $0.23^{+0.13}_{-0.20}$ \\
    NSM\_J4      & 5, 6    & $m^*_{0.8}$         &  100 (1038) & $0.253^{+0.098}_{-0.071}$ & $21.18^{+13.19}_{- 7.64}$  & $  4.39^{+ 3.43}_{- 1.90}~$ & $0.26^{+0.18}_{-0.21}$ \\
    NSM\_J5      & 5, 6    & $m^*_\mathrm{S}$    &  100 (1083) & $0.241^{+0.096}_{-0.080}$ & $19.56^{+ 8.33}_{- 5.63}~$ & $  4.73^{+ 3.49}_{- 1.93}~$ & $0.21^{+0.15}_{-0.16}$ \\
    NSM\_J6      & 5, 6    & Shen                &   100 (827) & $0.264^{+0.066}_{-0.054}$ & $20.61^{+ 6.71}_{- 4.77}~$ & $  5.26^{+ 3.78}_{- 3.02}~$ & $0.25^{+0.13}_{-0.19}$ \\
    NSM\_J7      & 5, 6    & SkShen              &   100 (908) & $0.270^{+0.094}_{-0.067}$ & $22.84^{+11.95}_{- 5.97}$  & $  5.34^{+ 3.93}_{- 2.97}~$ & $0.24^{+0.25}_{-0.19}$ \\
    NSM-DISK\_W1 & 7, 8    & S-def               &   100 (804) & $0.222^{+0.042}_{-0.038}$ & $15.59^{+ 6.73}_{- 4.45}~$ & $ 82.89^{+90.30}_{-42.84}$  & $0.41^{+0.30}_{-0.27}$ \\
    NSM-DISK\_W2 & 7, 8    & s6                  &   100 (533) & $0.197^{+0.052}_{-0.043}$ & $15.20^{+ 8.24}_{- 4.73}~$ & $ 82.73^{+80.53}_{-47.95}$  & $0.40^{+0.25}_{-0.25}$ \\
    NSM-DISK\_W3 & 7, 8    & $\alpha0.10$        &   100 (517) & $0.206^{+0.106}_{-0.091}$ & $14.47^{+ 8.83}_{- 5.38}~$ & $ 31.58^{+18.03}_{-12.02}$  & $0.21^{+0.10}_{-0.16}$ \\
    NSM-NDW\_M   & 9, 10   & -                   & 100 (17330) & $0.322^{+0.029}_{-0.023}$ & $14.28^{+ 1.65}_{- 1.31}~$ & $ 13.88^{+ 2.44}_{- 2.67}~$ & $0.09^{+0.01}_{-0.01}$ \\
    SN-NDW\_B1   & 11      & -                   &          36 & $0.434^{+0.026}_{-0.028}$ & $76.09^{+40.67}_{-26.07}$  & $  6.40^{+ 4.88}_{- 2.76}~$ & $0.08^{+0.08}_{-0.06}$ \\
    SN-NDW\_B2   & 12      & -                   &           4 & $0.450^{+0.035}_{-0.035}$ & $56.09^{+33.79}_{-21.84}$  & $  9.27^{+ 5.72}_{- 3.54}~$ & $0.03^{+0.01}_{-0.01}$ \\
    MRSN\_R1     & 13, 14  & 35OC-Rs: Fe         &           8 & $0.484^{+0.022}_{-0.021}$ & $13.58^{+ 6.89}_{- 3.82}~$ & $ 92.78^{+61.87}_{-18.73}$  & $0.16^{+0.09}_{-0.12}$ \\
    MRSN\_R2     & 13, 14  & 35OC-Rs: Fe-weak-r  &          10 & $0.394^{+0.038}_{-0.041}$ & $13.14^{+ 2.06}_{- 2.09}~$ & $104.80^{+24.84}_{-15.75}$  & $0.10^{+0.07}_{-0.07}$ \\
    MRSN\_R3     & 13, 14  & 35OC-Rs: r-process  &          10 & $0.242^{+0.016}_{-0.028}$ & $ 8.63^{+ 0.94}_{- 1.02}~$ & $ 89.24^{+24.05}_{-19.54}$  & $0.22^{+0.06}_{-0.05}$ \\
    MRSN\_R4     & 13, 14  & 35OC-Rs: weak-r     &           9 & $0.310^{+0.044}_{-0.049}$ & $11.59^{+ 2.66}_{- 1.68}~$ & $ 92.14^{+60.78}_{-24.61}$  & $0.11^{+0.02}_{-0.03}$ \\
    MRSN\_W      & 15      & -                   &           8 & $0.183^{+0.002}_{-0.003}$ & $ 6.22^{+ 0.36}_{- 0.27}~$ & $  6.76^{+ 1.36}_{- 1.51}~$ & $0.28^{+0.04}_{-0.06}$ \\
    \hline
    \end{tabular}
    \tablecomments{The model names include the following abbreviations: `NSBH' for neutron star black hole merger; `NSM' for binary neutron star merger; `SN' for normal core-collapse supernova; `MRSN' for magneto-rotational supernova. The suffixes `-DISK' and `-NDW' mark models that focus on the disk or neutrino-driven wind components.  Up to 100 tracers of each model have been selected to compare to our model - numbers in brackets indicate the total number of available tracers. For each model the mean value and standard deviation of all tracers are shown for the initial (i.e., $T=7\,\mathrm{GK}$) electron fraction, specific entropy, as well as for the expansion timescale. In addition, the average mass fraction difference between tracer and model prediction is given.}
    {\textsc{References}---(1)~\citet{Korobkin2012_AstrophysicalRobustnessNeutron}; (2)~\citet{Piran2013_ElectromagneticSignalsCompact}; (3)~\citet{Rosswog2013_MultimessengerPictureCompact}; (4)~\citet{Bovard2017_RprocessNucleosynthesisMatter}; (5)~\citet{Jacobi2023_EffectsNuclearMatter}; (6)~\citet{Ricigliano2024_ImpactNuclearMatter}; (7)~\citet{Fernandez2013_DelayedOutflowsBlack}; (8)~\citet{Wu2016_ProductionEntireRange};(9)~\citet{Martin2015_NeutrinoDrivenWindsAftermath}; (10)~\citet{Perego2014_NeutrinodrivenWindsNeutron}; (11)~\citet{Bliss2020_NuclearPhysicsUncertainties}\footnote{In the neutrino-driven wind model of (11) neutrinos play a critical role, e.g., by allowing the $\nu p$-process, whereas here we do not use neutrinos.}; (12)~\citet{Bliss2018_SurveyAstrophysicalConditions} (13)~\citet{Obergaulinger2017_ProtomagnetarBlackHole}; (14)~\citet{Reichert2021_NucleosynthesisMagnetorotationalSupernovae}; (15)~\citet{Winteler2012_MagnetorotationallyDrivenSupernovae}.}
\end{table*}

\begin{figure}
    \centering
    \includegraphics[width=\columnwidth]{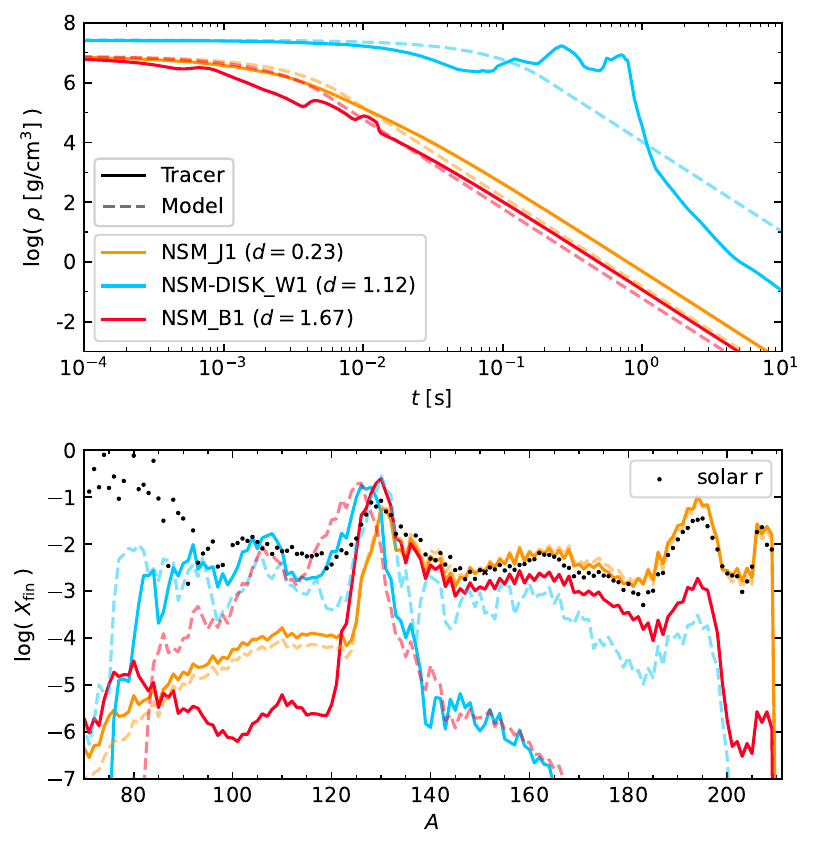}
    \caption{Density evolution and final mass fractions for three selected trajectories (solid lines) and their matching parametrizations (dashed lines). The tracer from NSM\_J1 shows a typical degree of agreement using the metric defined in Eq.~(\ref{eq:massFractionDistance}). The other two tracers result in a poor agreement with the model due to a non-monotonic density evolution with strong hydrodynamical shocks (NSM-DISK\_W1) and highly sensitive conditions on the border between two nucleosynthesis groups (NSM\_B1).}
    \label{fig:trajectoryComparison}
\end{figure}

Figure~\ref{fig:trajectoryComparison} shows a comparison of the density evolution and the final mass fractions for three tracers from hydrodynamic simulations (solid lines) and the closest parametric models (dashed lines). The selected tracer from model NSM\_J1 results in an agreement that is close to the mean value of all considered trajectories. The other two cases correspond to the few outliers. In one case (NSM-DISK\_W1), the differences are caused by a non-monotonic density behavior that cannot be reproduced by the parametric description. In the other case (NSM\_B1), the trajectory has $Y_\mathrm{e,0} = 0.27, s_\mathrm{0} = 21.04\,\mathrm{k_B/nuc}, \tau = 1.79\,\mathrm{ms}$ and, therefore, lies in the transition region between the two abundance groups G2 and G4, cf. Fig.~\ref{fig:clustersByPeaks}. Even if the density evolution is reproduced well by the parametric trajectory, small differences in the hydrodynamic quantities have a large impact on the nucleosynthesis results. In these conditions, post-processing tracers from simulations is particularly important to get detailed nucleosynthesis yields.

\section{Constraints from observations}
\label{sec:comparisonToObservations}

In this section, we compare the final abundances of our model with the solar r-process residual abundances from \cite{Sneden2008_NeutronCaptureElementsEarly}, which contain complete isotopic abundance data. However, they are prone to uncertainties from s-process modeling and are a mix of multiple r-process events \citep{Goriely1999_UncertaintiesSolarSystem}. Therefore,  we also compare elemental abundances to metal-poor r-process enhanced stars as those exhibit the fingerprint of one or few r-process events.

\subsection{Second and third peaks}
\label{sec:peakLocations}

\begin{figure*}
    \centering
    \includegraphics[height=0.395\textwidth]{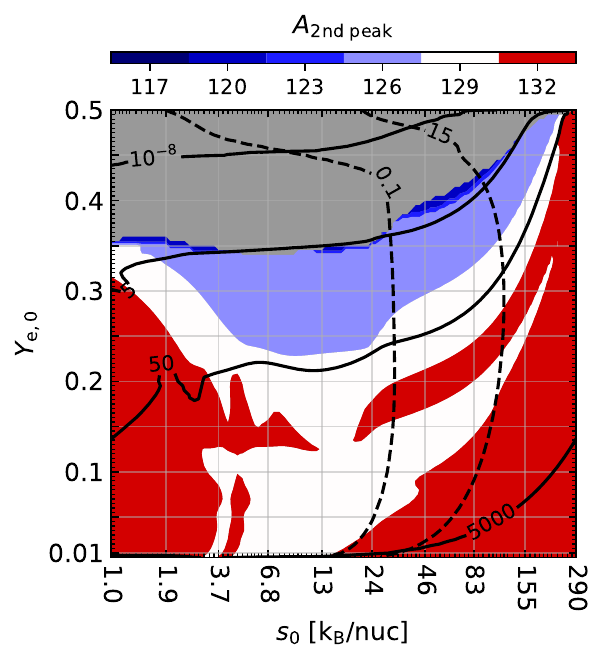}
    \hfill
    \centering
    \includegraphics[height=0.395\textwidth]{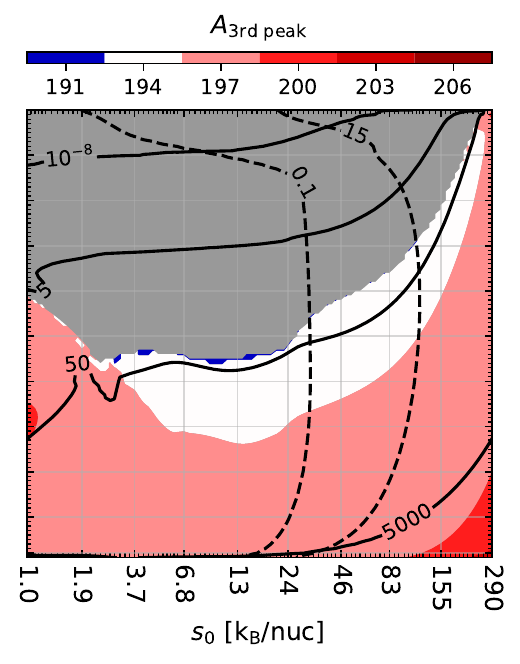}
    \hfill
    \centering
    \includegraphics[height=0.395\textwidth]{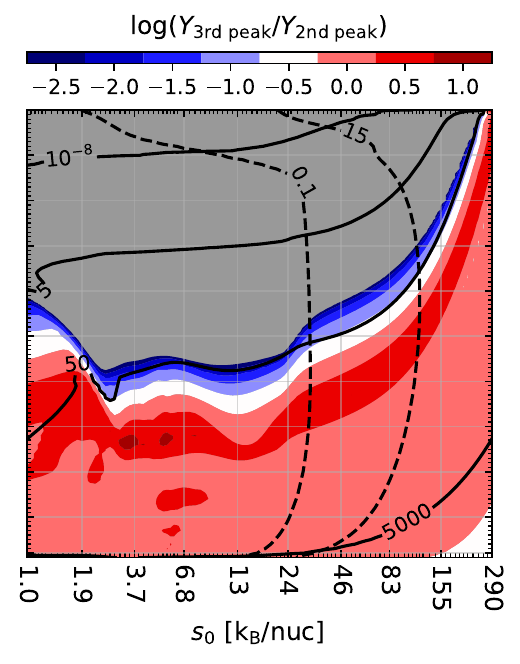}
    \hfill
    \caption{\textbf{Left:} Center mass number of second r-process peak as obtained by a Gaussian fit. The agreement with solar is best for the white regions, while in the blue (red) areas the peak shifts to lower (higher) mass numbers. \textbf{Center:} Same for the third r-process peak. \textbf{Right:} Same but for the ratio of the abundance heights of third and second peaks. All panels show the results for $\tau=8\,\mathrm{ms}$.}
    \label{fig:peakLocation}
\end{figure*}

The exact location of the r-process peaks can vary depending on the astrophysical conditions and also the nuclear physics \citep[see e.g.,][]{Surman1997_SourceRareEarthElement,Meyer1997_SurveyRProcessModels,Arcones2011_DynamicalRprocessStudies,Mumpower2017_ReverseEngineeringNuclear}. In order to investigate the deviation of the calculated peaks compared to solar, we fit Gaussian peaks to the $A$-intervals of the second and third peaks for all conditions. The left and center panels of Fig.~\ref{fig:peakLocation} show the resulting center values of the peaks for $\tau=8\,\mathrm{ms}$.

The location of the second peak agrees with solar ($A_\mathrm{2nd\,peak}\approx129$) for conditions with $Y_\mathrm{n}/Y_\mathrm{seed}\approx50$. In addition, some conditions with higher $Y_\mathrm{n}/Y_\mathrm{seed}$ also produce the second peak at the right location, but only for $3\,\mathrm{k_B/nuc}\lesssim s_\mathrm{0}\lesssim 20\,\mathrm{k_B/nuc}$, when the seed nuclei are not too heavy (cf. Fig.~\ref{fig:AbarHeavies3GK}). Otherwise, fission cycling can shift the peak towards heavier mass numbers. Note that this result is sensitive to the fission yield distributions.  On the other hand, for $Y_\mathrm{n}/Y_\mathrm{seed}<50$ the second peak shifts towards lower mass numbers, as the neutron flux is small and often does not significantly overcome N=82.

The location of the third r-process peak best agrees with solar ($A_\mathrm{3rd\,peak}\approx194$) in a similar range of conditions around $50\lesssim Y_\mathrm{n}/Y_\mathrm{seed}\lesssim 150$. For higher neutron-to-seed ratios, many neutrons are available after freeze-out and neutron captures on the third peak nuclei shift the peak towards heavier mass numbers. However, the neutron capture cross sections are still very uncertain \citep{Mumpower2016_ImpactIndividualNuclear} and if they may occur in other regions, the described shift of the third peak may be prevented \citep{Arcones2011_DynamicalRprocessStudies}. For extremely high neutron-to-seed-ratios (i.e., very high entropies), this shift can be as much as 10 mass numbers.

Similar trends for the location of the second and third peaks are found for most of the investigated expansion timescales. However, for very fast ($\tau\lesssim1\,\mathrm{ms}$) and very slow expansions ($\tau\gtrsim150\,\mathrm{ms}$), the two peaks shift towards higher mass numbers for most initial electron fractions and entropies, in agreement with the results in \citet{Meyer1997_SurveyRProcessModels}. In case of very fast expansions, this is due to the rapid temperature decrease and thus the freeze-out of $(n,\gamma)-(\gamma,n)-$equilibrium within few milliseconds. During the subsequent beta decay phase towards stability, there are still many neutrons available, leading to the shift of the peaks towards larger mass numbers. On the other hand, when the expansion is very slow, the $(n,\gamma)-(\gamma,n)-$equilibrium can persist for up to 10\,s and the r-process path moves closer to stability via beta decays. Therefore, the final peak locations also end up at heavier mass numbers than in solar.

In addition to the location of the r-process peaks, their relative heights are important. The right panel of Fig.~\ref{fig:peakLocation} shows the ratio of the third and second peaks. The solar value ($\log[Y_\mathrm{3rd\,peak}/Y_\mathrm{2nd\,peak}]\approx-0.5$) is only reached for a small part of the parameter space with $Y_\mathrm{e}\approx0.175$, where again $Y_\mathrm{n}/Y_\mathrm{seed}\approx50$. For higher neutron-to-seed ratios, the third peak is overproduced. However, there are at least two uncertainties that may affect these results. First, the third-to-second peak ratio is maximum for $Y_\mathrm{n}/Y_\mathrm{seed}\approx150$, where significant amounts of third-peak elements and fission nuclei are produced but the fission cycle is not completed. The fission yields from \citet{Mumpower2020_PrimaryFissionFragment} used here present a broad distribution in mass number. Narrower distributions, e.g., the more approximate scheme in \citet{Panov2005_CalculationsFissionRates}, can produce higher abundances of second peak nuclei through fission, and thus decrease the ratio of third to second peaks in conditions where fission occurs. Second, the second peak is also produced for different conditions without being co-produced with the third peak. Therefore, an observed r-process pattern could include the contribution of several different conditions for the region of the second peak (see Sect.~\ref{sec:solarComparison2Components}).

Note that the conditions for which the peak ratio agrees with solar largely overlap with the conditions where the solar locations of the second and third peak are reproduced. In addition, these conditions are found in many hydrodynamical simulations, cf. Fig.~\ref{fig:YeSTauMapping}.

\subsection{How many r-processes from 1st to 3rd peak?}
\label{sec:solarComparison2Components}

\begin{figure}
    \centering
    \includegraphics[width=\columnwidth]{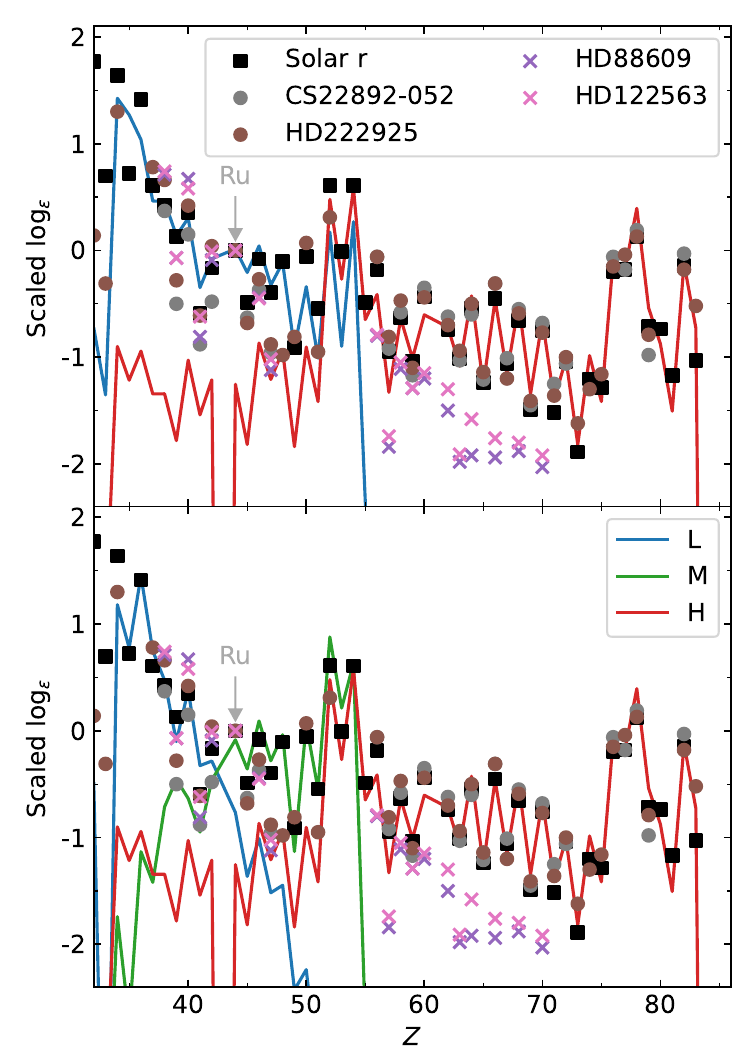}
    \caption{Symbols show stellar abundances, scaled to Ru ($Z=44$): Solar r residuals (squares), main r-process stars (dots) from \citet{Sneden2003_ExtremelyMetalpoorNeutron,Roederer2022_RprocessAllianceNearly}, limited r-process stars (crosses) from \citet{Honda2007_NeutronCaptureElementsVery}.  Solid lines show model abundances scaled to solar r. \textbf{Top:} only two components are used to fit the abundances: L-component (blue, $Y_\mathrm{{e,0}}=0.325,~s_\mathrm{{0}}=1.1\,\mathrm{{k_B/nuc}},~\tau=16\,\mathrm{{ms}}$) and H-component (red, $Y_\mathrm{{e,0}}=0.165,~s_\mathrm{{0}}=12.2\,\mathrm{{k_B/nuc}},~\tau=2\,\mathrm{{ms}}$). \textbf{Bottom:} three components are used to fit the abundances: L-component (blue, $Y_\mathrm{{e,0}}=0.355,~s_\mathrm{{0}}=1.2\,\mathrm{{k_B/nuc}},~\tau=2\,\mathrm{{ms}}$), M-component (green, $Y_\mathrm{{e,0}}=0.300,~s_\mathrm{{0}}=83.0\,\mathrm{{k_B/nuc}},~\tau=150\,\mathrm{{ms}}$), and H-component (same as in upper panel).}
    \label{fig:stellarComparison}
\end{figure}

\begin{figure}
    \centering
    \includegraphics[width=\columnwidth]{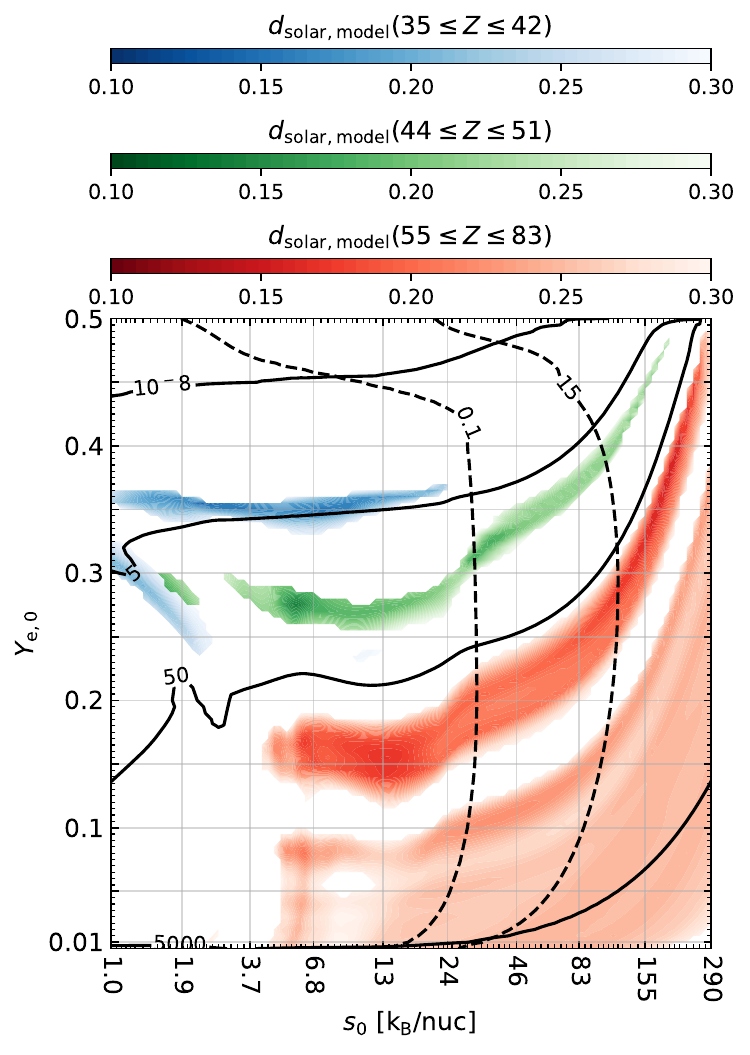}
    \caption{Conditions that best produce the solar abundances of the L- (blue), M- (green) and H-component (red) for $\tau=8\,\mathrm{ms}$.}
    \label{fig:solarAgreement3Components}
\end{figure}

As discussed in Sect.~\ref{sec:groups}, the first and third peak elements are rarely produced with a single condition and even in the few cases of co-production the relative abundances of first to third peak are too low compared to solar. Therefore, a superposition of conditions is required to explain the full r-process pattern. Such conditions could come from different events or be co-produced in a single event in various components of the ejecta. For example, in neutron star mergers some conditions could be found in the dynamical ejecta and others in the disk ejecta or neutrino-driven wind. Following \cite{Hansen2014_HowManyNucleosynthesis}, we first separate the full r-process into two components\footnote{The second peak elements ($52\leq Z \leq 54$) are excluded from the two intervals because they can be produced in both components.}: `L-component': $35\leq Z \leq 51$\footnote{Our conclusions are not affected by choosing a smaller interval only up to Z=49 or 50.} and `H-component': $55\leq Z \leq 83$. The upper panel of Fig.~\ref{fig:stellarComparison} shows the combination of two components based on two sets of conditions from our parametric model that best reproduce the solar abundances\footnote{We find similar result if we use in stead of solar abundances those of CS22892-052 \cite{Sneden2003_ExtremelyMetalpoorNeutron} and CS222925 \cite{Roederer2022_RprocessAllianceNearly}, considered as typical r-process patterns.}. In addition to the abundances of solar r-process (squares), four metal-poor stars with main r-process (dots) and limited r-process (crosses) patterns are shown.

We have explored the regions of the parameter space where the L- and H-components are produced and there is no overlap. The H-component as in solar is produced for a wide range of conditions with $Y_\mathrm{n}/Y_\mathrm{seed}\gtrsim50$, similar to Fig.~\ref{fig:peakLocation} and Sect.~\ref{sec:peakLocations}. This again confirms that the observed robustness of the r-process pattern between the second and third peaks may be due to a wide range of conditions producing similar abundances. For the L-component a good agreement to solar is found only for few conditions: For medium to slow expansions ($\tau\gtrsim8\,\mathrm{ms}$), low to medium entropies ($1\,\mathrm{k_B/nuc}\lesssim s_\mathrm{0}\lesssim50\,\mathrm{k_B/nuc}$), and a narrow electron fraction interval around $Y_\mathrm{e}\approx0.3$. Some conditions with very short expansion timescales and extremely high entropy also produce solar-like abundances. 

The few conditions that produce the L-component are not found in current hydrodynamical simulations. This might hint towards a third `M-component', which produces predominantly the elements between $41\leq Z\leq51$. It has been shown before \cite[see e.g.,][]{Hansen2011_OriginPalladiumSilver} that this might be required to explain the observed variability in stellar abundances between Sr, Y, Zr, and the heavier elements around Ag. The elements of the M-component may result from fission as suggested by \citet{Roederer2023_ElementAbundancePatterns}. We also find conditions in which the M-component is produced via fission. However, such conditions always overproduce the rare earth region, due to the roughly symmetric fission yields used here \citep{Mumpower2020_PrimaryFissionFragment}. 

The best fits of model abundances to solar r-process for each of the three components are shown in the lower panel of Fig.~\ref{fig:stellarComparison}, with relative weights of approximately 50\%, 25\%, and 25\% for the L-, M-, and H-components. The regions in the parameter space that produce each component are marked with the same colors in Fig.~\ref{fig:solarAgreement3Components} for $\tau=8\,\mathrm{ms}$, similar results are found for the other timescales. Notice that this L-component corresponds to the case of using three components. When only two are used, this region is smaller and not reached by simulations. For calculating these regions, we use the following version of the distance metric introduced in Eq.~(\ref{eq:massFractionDistance}):
\begin{align}
    \label{eq:elementalAbundancesDistance}
    d_\mathrm{solar,model}=\frac{1}{\Delta Z}&\sum_{Z=Z_\mathrm{min}}^{Z_\mathrm{max}}\Big|\Big(\log \big(Y_\mathrm{solar}(Z)\big) - \bar{Y}^\mathrm{log}_\mathrm{solar} \Big) - \notag \\
    &\Big(\log\big(Y_\mathrm{model}(Z)\big) - \bar{Y}^\mathrm{log}_\mathrm{model} \Big)\Big|\,,
\end{align}
where $\Delta Z=Z_\mathrm{max}-Z_\mathrm{min}+1$ is used for normalization. The terms $\bar{Y}^\mathrm{log}_\mathrm{solar/model}$ are the average logarithmic abundances of solar/model in the considered $Z$--interval and ensure the appropriate scaling. Some conditions lead to a significant overproduction of elements outside the considered $Z$--interval and are therefore excluded in our analysis.

This preliminary comparison to observation aims to show the potential of our model and will be extended in future work. Still, we can already draw important implications for the r-process, even if some of them have already been discussed in other contexts, here we proved them in a more systematic way. First, a combination of at least three components is necessary to explain the full r-process pattern with conditions found in current simulations. Such a combination could come from adding the contributions of different parts of the ejecta in one single event or from mixing ejecta from different events. Second, the region between the second and third peaks, i.e., the H-component, is robustly produced for a large region of the parameter space. Third, observed r-process abundances strongly constrain the mixing of the different components, showing a puzzling \emph{second robustness}: the relative abundance between the L- (or L+M, in the case of three components) and the H-component is the same in the solar r-process and in the many main r-process stars observed (e.g., CS22892-052 and HD222925 shown in Fig.~\ref{fig:stellarComparison}). Still variations are found for r-limited stars, showing also the possibility of different mixing of the two (or three) components. This variation between the L- and H-components is also relatively small in r-limited stars. However, one may argue that very low abundances of the H-component cannot be observed. Nevertheless, there are no observations of stars with only an L-component (i.e., r-free stars without Eu or Ba).

\section{Summary and conclusions}
\label{sec:conclusions}

In this study, we presented state-of-the-art r-process network calculations for 120\,000 astrophysical conditions that differ in their initial electron fraction $Y_\mathrm{e,0}$, initial entropy $s_\mathrm{0}$, and expansion timescale $\tau$. These conditions include and extend beyond the conditions found in current hydrodynamical simulations of proposed r-process scenarios, including binary neutron star mergers, black hole neutron star mergers, their disks, magneto-rotational supernovae, and neutrino-driven winds. In general, we find good agreement between our model and nucleosynthesis results based on tracers from simulations. Significant discrepancies are only found for few trajectories, either when they have highly non-monotonic density profiles (e.g., hydrodynamical shocks) or when they are close to the boundary condition that determines the production of heavy elements beyond the second peak.

Our results confirm many previous works \cite[see e.g.,][]{Hoffman1997_ModelIndependentRprocess,Hoffman1997_NucleosynthesisNeutrinodrivenWinds,Meyer1997_SurveyRProcessModels,Freiburghaus1999_AstrophysicalRprocessComparison,Arnould2007_RprocessStellarNucleosynthesis,Lippuner2015_RprocessLanthanideProduction} in that low electron fractions, high entropies, and fast expansion timescales facilitate a strong r-process, since they lead to a high neutron-to-seed ratio at $3\,\mathrm{GK}$, when the rapid neutron capture phase approximately starts. In addition, the average mass number of seed nuclei $\bar{A}_\mathrm{seed}$ at $3\,\mathrm{GK}$ is a crucial quantity that sets the starting conditions on which neutron captures can operate. We find large $\bar{A}_\mathrm{seed}\approx100-120$ in two cases: (1) The ion-dominated regime, in which heavy nuclei have formed already during NSE and (2) the parameter region in which the alpha-to-seed ratio after charged particle freeze-out is around one.

All conditions can be classified into eight nucleosynthesis groups ranging from only iron-group production to full r-process with fission cycling. The full r-process pattern with all three peaks is only found in 1.4\% of the conditions. For neutron-to-seed ratios around 50 the second and third r-process peaks form without actinides. Higher neutron-to-seed ratios lead to multiple fission cycles and a robust production of the second and third peak nuclei, including actinides. This group is very common (36.8\% of the parameter space) indicating that the robustness between the second and third peaks may be explained by the fact that many different conditions produce the same abundance pattern.

We showcased how our model can be compared to observed abundance patterns, although future works will elaborate on this. The location and relative heights of the second and third solar r-process peaks can be reproduced for neutron-to-seed ratios around 50. Higher neutron-to-seed ratios can shift the peaks towards heavier mass numbers and can result in a relative overproduction of the third peak, although these results strongly depend on the nuclear physics input.
Since no conditions were found to produce all three r-process peaks at once, combinations of two and three conditions were tested. The elements between the second and third peaks (H-component) are produced in agreement with solar for a large part of the parameter space. The first to second peak elements (L-component), on the other hand, only agree for limited regions, which are not commonly found in simulations. When they are separated into two components again (L- and M-components), they can be produced more robustly and for conditions found in current simulations.

\begin{acknowledgments}
We thank Camilla J. Hansen, Thanassis Psaltis, and Friedel Thielemann for useful discussions. This work was supported by the Deutsche Forschungsgemeinschaft (DFG, German Research Foundation) – Project-ID 279384907 – SFB 1245 and the State of Hessen within the Research Cluster ELEMENTS (Project ID 500/10.006). M.R. acknowledges support from the grants FJC2021-046688-I and PID2021-127495NB-I00, funded by MCIN/AEI/10.13039/501100011033 and by the European Union "NextGenerationEU" as well as "ESF Investing in your future". Additionally, he acknowledges support from the Astrophysics and High Energy Physics program of the Generalitat Valenciana ASFAE/2022/026 funded by MCIN and the European Union NextGenerationEU (PRTR-C17.I1). This publication benefited highly from collaborations and exchange within the European Union’s Horizon 2020 research and innovation program under grant agreement No. 101008324 (ChETEC-INFRA) and the "International Research Network for Nuclear Astrophysics" (IReNA). 
\end{acknowledgments}

\bibliography{bibliography}{}
\bibliographystyle{aasjournal}

\end{document}